\documentclass[useAMS]{mn2e}
\usepackage{graphicx}
\newcommand{\gapprox}{\,\rlap{\lower 2.5pt 
\hbox{$\sim$}}\raise 1.5pt\hbox{$>$}\,}
\newcommand{\gsim}{\,\rlap{\lower 2.5pt 
\hbox{$\sim$}}\raise 1.5pt\hbox{$>$}\,}
\newcommand{\lapprox}{\,\rlap{\lower 2.5pt 
\hbox{$\sim$}}\raise 1.5pt\hbox{$<$}\,}
\newcommand{\lsim}{\,\rlap{\lower 2.5pt 
\hbox{$\sim$}}\raise 1.5pt\hbox{$<$}\,}
\newcommand{\msun}{{M_{\odot}}}

\newcommand{\chiquadr}{\ensuremath{\chi^{2}_{\rm r}}}

\newcommand{\zform}{\ensuremath{z_{\rm_f}~}}

\def\eeq{\end{equation}}
\def\beq{\begin{equation}}

\title[The star formation rates and masses of $z\sim 2$ galaxies]{Star formation rates and masses of $z\sim 2$ galaxies from multicolour photometry}
\author[C. Maraston et al.]{Claudia Maraston$^{1}$\thanks{E-mail: claudia.maraston@port.ac.uk}, Janine Pforr$^{1}$, Alvio Renzini$^{2}$, Emanuele Daddi$^3$, \newauthor  Mark Dickinson$^{4}$,  Andrea Cimatti$^5$, Chiara Tonini$^1$\\
$^{1}$Institute of Cosmology and Gravitation, University of Portsmouth, Dennis Sciama Building, Burnaby Road, PO1 3FX Portsmouth, UK\\
$^{2}$INAF-Osservatorio Astronomico di Padova,Vicolo dell'Osservatorio 5, I-35122 Padova, Italy\\
$^{3}$CEA,  Irfu/SAp, F-91191 Gif-sur-Yvette, France, France\\
$^{4}$National Optical Astronomical Observatory, 950 N. Cherry Avenue, Tucson, AZ 85719, USA \\
$^{5}$Dipartimento di Astronomia, Universit\`a  di Bologna, Via Ranzani, I-40126 Bologna, Italy\\}
\begin{document}
\date{in press}
%
\pagerange{\pageref{firstpage}--\pageref{lastpage}} \pubyear{2010}
\maketitle
\label{firstpage}
\begin{abstract}
 Fitting synthetic spectral energy distributions (SED) to the
 multi-band photometry of galaxies to derive their star formation
 rates (SFR), stellar masses, ages, etc. requires making a priori
 assumptions about their star formation histories (SFH). A widely
 adopted parameterization of the SFH, the so-called $\tau$-models
 where SFR $\propto {\rm e}^{-t/\tau}$ is shown to lead to
 unrealistically low ages when applied to a sample of actively star
 forming galaxies at $z\sim 2$, a problem shared by other SFHs
 when the age is left as a free parameter in the fitting
 procedure. This happens because the SED of such galaxies, at all
 wavelengths, is dominated by their youngest stellar populations,
 which outshine the older ones. Thus, the SED of such galaxies conveys
 little information on the beginning of star formation, i.e., on the
 age of their oldest stellar populations.  To cope with this problem,
 besides $\tau$-models (hereafter called direct-$\tau$ models), we
 explore a variety of SFHs, such as constant SFR and inverted-$\tau$
 models (with SFR $\propto {\rm e}^{+t/\tau}$), along with various
 priors on age, including assuming that star formation started at high
 redshift in all the galaxies in the test sample. We find that
 inverted-$\tau$ models with such latter assumption give SFRs and
 extinctions in excellent agreement with the values derived using only
 the UV part of the SED, which is the one most sensitive to
 ongoing star formation and reddening.  These models are also shown to
 accurately recover the SFRs and masses of mock galaxies at $z\sim 2$
 constructed from semi-analytic models, which we use as a further
 test.  All other explored SFH templates do not fulfil these two test
 as well as inverted-$\tau$ models do. In particular, direct-$\tau$
 models with unconstrained age in the fitting procedure overstimate
 SFRs and underestimate stellar mass, and would exacerbate an apparent
 mismatch between the cosmic evolution of the volume densities of SFR
 and stellar mass.  We conclude that for high-redshift star forming
 galaxies an exponentially increasing SFR with a high formation
 redshift is preferable to other forms of the SFH so far adopted in
 the literature.
\end{abstract}
\begin{keywords}
galaxies: evolution --- galaxies: starbursts --- galaxies: high-redshift
\end{keywords}
\section{Introduction}
The evolution of the baryonic component in galaxies is the hardest
part to model in galaxy formation theories. How gas is accreted onto
dark matter halos, its thermal history, how it is turned into stars,
and if, how and when such star formation is quenched cannot be reliably 
predicted from first principles. The physical processes that are involved are
too complex and non linear, with hydrodynamic simulations failing by a
large margin to cover the extremely wide dynamical range that would be
required to describe phenomena ranging from the formation of stars and
supermassive black holes to the behaviour of multiphase gas on
megaparsec scales. Where  it is hard to progress with pure theory alone,
observations can help and lead to further advances in our understanding
of how galaxies form and evolve. Indeed, over the last two decades a
wealth of multiwavelength observations from both ground and space
facilities have provided us with a rich vision of the galaxy
populations in the high redshift universe. Once correctly
interpreted, such photometric, spectroscopic, and high spatial
resolution data can provide us with estimates of stellar mass ($M_*$),
star formation rate (SFR) and star formation history (SFH), structure,
dynamics and nuclear activity, for a very large number of galaxies,
and do so as a function of redshift and environment.

In particular, the  accurate measurement of stellar mass
and SFR is critical for trying to establish the evolutionary links
connecting galaxy populations at one redshift with those at another
redshift. Setting constraints on the previous SFH of individual
galaxies is also relevant in this context. All this is currently
obtained by fitting the spectral energy distribution (SED) of
synthetic stellar populations to the SED of galaxies from their multi-band
photometry. In practice, a large set of template synthetic
SEDs is constructed with different SFHs, extinctions and
metallicities, and a best fit is sought by picking the SFH that minimises
the $\chiquadr$. 

In this paper we focus on star forming galaxies at $1.4\lapprox
z\lapprox 2.5$, hence covering the epoch of major star formation
activity, and explore a wide set of SFHs showing advantages and
disadvantages of different parameterizations of them. The case of
passively evolving galaxies at $z\gapprox 1.4$ was already addressed in
Maraston et al. (2006).

It is well known that at $z\sim 2$ galaxies with SFRs as high as a few
$\sim 100 M_{\odot}/{\rm yr}^{-1}$ are quite common (e.g., Daddi et
al. 2005), and by analogy with the rare objects at $z\simeq 0$ with
similar SFRs (like Ultra Luminous Infra-Red Galaxies, ULIRGs), it was
widely believed that such galaxies were caught in a merging-driven
starburst. However, integral-field near-infrared spectroscopy has
revealed that at least some of these galaxies have ordered, rotating
velocity fields with no kinematic evidence for ongoing merging (Genzel
et al.\ 2006). Still, the disk was shown to harbour several
star-forming clumps and to have high velocity dispersion and gas
fraction. All this makes such disk quite different from local disk
galaxies, and it is now well documented that the same properties apply
to many similar objects at $z\sim 2$ (F\"orster-Schreiber et
al. 2009).

\begin{figure}
\includegraphics[width=0.49\textwidth]{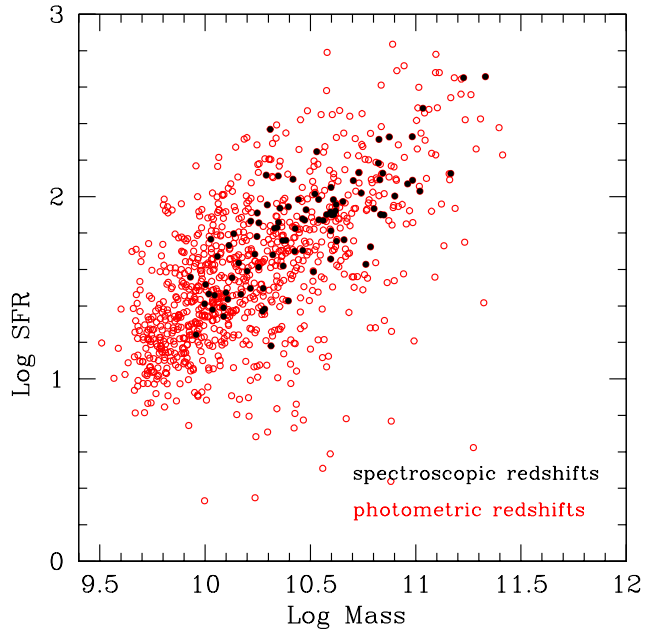}
\caption{The SFR vs. stellar mass of star forming $BzK$-selected
galaxies in the GOODS-South field (from Daddi et al. 2007a). The
sub-sample with spectroscopic redshifts is indicated, and represents
the set of galaxies for which various SED fits are attempted in 
this paper. The SFRs are in $\msun$/yr and masses in $\msun$ units.
\label{sfrm}}
\end{figure}

That high SFRs in $z\sim 2$ galaxies do not necessarily imply {\it
starburst} activity became clear from a study of galaxies in the GOODS
fields (Daddi et al. 2007a). Indeed, for starforming galaxies at
$1.4\lapprox z\lapprox 2.5$ the SFR tightly correlates with stellar
mass (with SFR $\propto\;\sim M_*$), with small dispersion ($\sim 0.3$
dex), as shown in Figure~\ref{sfrm}. Only a few galaxies lie far away
from the correlation: a relatively small number of passive galaxies
(with undetectable SFR, not shown in the figure), and sub-mm galaxies
(SMG) with much higher SFRs, which may indeed be the result of
gas-rich major mergers.  Among starforming galaxies, the small
dispersion of the SFR for given $M_*$ demonstrates that these objects
cannot have been caught in a special, starburst moment of their
existence.  Rather, they must sustain such high SFRs for a major
fraction of the time interval between $z=2.5$ and $z=1.4$, i.e. for
some $1-2$ Gyr, instead of the order of one dynamical time
($\sim 10^8$ yr) typical of starbursts. Similar correlations have also
been found at lower redshifts, notably at $z\sim 1$ (Elbaz et
al. 2007), $0.2\sim z\sim 1$ (Noeske et al.  2007), and $z\sim 0$
(Brinchmann et al. 2004).

In a recent study, Pannella et al. (2009) have measured the average
SFR vs. stellar mass for $1.4\lsim z\lsim 2.5$ galaxies using the 1.4
GHz flux from the VLA coverage of the COSMOS field (Schinnerer et al.
2007). When combined also with data at lower redshifts (see references
above), Pannella et al. derive the following relation for the average
SFR as a function of galaxy mass and time:
\begin{equation}
<{\rm SFR}>\simeq 270\times (M_*/10^{11}\msun)\times (t/3.4\times 
10^9 {\rm yr})^{-2.5},
\end{equation}
where the SFR is in $\msun$/yr units and $t$ is the cosmic time, i.e.,
the time since the Big Bang. Beyond $z\sim 2.5$ ($t\lsim 2.7$
  Gyr) the specific SFR (= SFR/$M_*$) appears to flatten out and
  remain constant all the way to very high redshifts (Daddi et
  al. 2009; Stark et al. 2009; Gonzalez et al. 2009).

In parallel with these observational evidences, theorists are shifting
their interests from (major) mergers as the main mechanism to grow
galaxies, to continuous {\it cold stream} accretion of baryons, that
are then turned into stars in a quasi-steady fashion (e.g., Dekel et
al. 2009). Clearly, a continuous, albeit fluctuating SFR such as in
these cold stream models provides a far better match to the observed tight
SFR$-M_*$ relation, compared to a scenario in which star formation
proceeds through a series of short starbursts interleaved by long
periods of reduced activity. This is not to say that major mergers do
not play a role. They certainly exist, and can lead to real giant
starbursts bringing galaxies to SFRs as high as $\sim 1000\;
M_{\odot}/{\rm yr}$, currently identified with SMGs (e.g. Tacconi et
al. 2008; Menendez-Delmestre et al. 2009).

Deriving SFRs, ages, stellar masses, etc. from SED fitting requires
making assumptions on the previous SFH of galaxies. A
widespread approach is to fit the SED of galaxies at low as well as high
redshifts with so called ``$\tau$-models", i.e., synthetic SEDs in
which the SFH is described by an exponentially declining SFR, starting
at cosmic time $t_\circ$, i.e.:
\begin{equation}
{\rm SFR}=A{\rm e}^{-(t-t_\circ)/\tau},
\end{equation}
where $A=$SFR($t=t_\circ$).
For a galaxy at cosmic time $t$ a $\chi^2$ fit then gives the age
(i.e., $t-t_\circ$, the time elapsed since the beginning of star
formation), the SFR e-folding time $\tau$, the reddening $E(B-V)$, the
metallicity $Z$ and finally the stellar mass $M_*$ via the scale factor
$A$. The SFR then follows from Equation (2), where $t$ is the cosmic time
corresponding to the observed redshift of each galaxy. Of course, the
reliability of the results depends on the extent to which the actual
SFH is well represented by a declining exponential.

It is worth recalling that the first applications of $\tau$-models were
to figure out the ages of local elliptical galaxies, and the typical
result was that the age is of the order of one Hubble time, and $\tau$
of the order of 1 Gyr or less (e.g., Bruzual 1983). This approach
confirmed that the bulk of stars in local ellipticals are very old,
hence formed within a short time interval compared to the present
Hubble time. Later, the usage of $\tau$-models was widely extended
also to actively star forming galaxies at virtually all redshifts, to
the extent that it became the default assumption in this kind of
studies (e.g., Papovich et al. 2001; Shapley et al. 2005; Lee et
al. 2009; Pozzetti et al. 2009; F\"oster-Schreiber et al.  2009; Wuyts
et al. 2009b).  Such an assumed SFH may give reasonable results for
local spirals, as their SF activity has been secularly declining for
an order of one Hubble time (e.g., Kennicutt 1986), but we shall argue
that it may be a rather poor representation of the SFH of high-$z$
galaxies, and may lead to quite unphysical results.

Cimatti et al. (2008) noted that the age of elliptical galaxies at
$z\sim 1.6$ turns out to be $\sim 1$ Gyr both when using only the
rest-frame UV part of the SED, and when using the whole optical-to-near-IR SED in
conjunction with $\tau$-models.  However, they also noted that the
former ``age'' is actually the age of the population formed in the
last significant episode of star formation, while the latter ``age''
corresponds to the time elapsed since the beginning of star
formation. The near equality of these two ages suggests that the SFR
peaked shortly before being quenched, rather than having peaked at an
earlier time and having declined ever since. Moreover, using $\tau$
models one implicitly assumes that galaxies are all caught at their
minimum SFR, which is possibly justified for local ellipticals and
spirals, but not necessarily for star-forming galaxies at
high redshifts that may actually be at the peak of their SF activity.

Indeed, an integration of $dM_*/dt = <{\rm SFR}>$ where $ <{\rm SFR}>$
is given by Equation (1) shows that the SFR can {\it increase}
quasi-exponentially with time before the effect of the declining term
$t^{-2.5}$ takes over, or star formation is suddenly quenched and
the galaxy turns passive (Renzini 2009). Mass and SFR formally increase
exponentially when SFR $\propto\sim M_*$, independent of time, as 
appears to be the case for $z\gsim 2.5$ (Gonzales et al. 2010). Thus,
the observations of both passive and star-forming galaxies at
$1.4\lapprox z\lapprox 2.5$ suggest that the SFRs of these galaxies
may well have increased with time, rather than decreased. For these
reasons, in this paper we use both {\it direct}-$\tau$ models, with
the SFR given by Equation (2), as well as {\it inverted}-$\tau$ models
in which the SFR increases exponentially with time, i.e.:
\begin{equation}
{\rm SFR}=Ae^{+(t-t_\circ)/\tau}.
\end{equation}

Thus, direct-$\tau$ models assume that galaxies are caught at their
minimum SFR and had their maximum SFR at the beginning, whereas
inverted-$\tau$ models assume that galaxies are caught at their
maximum SFR and had their minimum SFR at the beginning. These two
extreme assumptions may to some extent {\it bracket} the actual SFHs of
real galaxies,  or at least of the majority of them which, because of the 
tight SFR$-M_*$ relation, must have a relatively smooth SFH.
Here we explore which of the two assumptions gives the 
better fit to the SED of $z\sim 2$ galaxies, and discuss the 
astrophysical plausibility of the relative results. Besides these
exponential SFHs we also consider the case of constant SFRs. In
principle, other, differently motivated SFHs could also be explored,
but in this paper we restrict the comparison to these three simple
options, with SFRs increasing with time, decreasing, or constant.

The paper is organised as follows. Section 2 provides information on
the galaxy data base that is used for our SED fitting
experiments. Section 3 illustrates the procedure of SED fitting and
describes the models. Section 4 presents the results we obtain for the
different star formation history templates and their
comparisons. Section 5 is devoted to a general discussion and
presents our conclusions.

Finally, we adopt a cosmology with $\Omega_{\Lambda}$, $\Omega_{M}$
and \\$h~=H_{0}/(100~{\rm km s^{-1} Mpc^{-1}})$ equal to 0.7, 0.3 and
0.75, respectively, for consistency with most previous works. The age of
the best-fit model is required to be lower than the age of the
Universe at the given spectroscopic redshift. The SFR and masses are always
in $\msun$/yr and $\msun$ units, respectively.
\section{Galaxy data}
We use a sample of 96 galaxies in the GOODS-South field from Daddi et
al. (2007a). These galaxies were selected via the $BzK$~diagram (Daddi
et al. 2004) to be star forming, as confirmed by the detection in deep
Spitzer+MIPS data at $24\mu$m for $>90$\% of the galaxies in the
sample. We include only objects with accurately determined
spectroscopic redshifts at $z>1.4$, which were derived from a variety
of surveys (see Daddi et al. 2007a for references), including notably
ultra-deep spectroscopy from the GMASS
project\footnote{http://www.arcetri.astro.it/~cimatti/gmass/gmass.html}(J. Kurk
et al. 2009, in preparation) and from the GOODS survey at the Very
large Telescope (Vanzella et al. 2008; Popesso et al. 2009). Galaxies
in the resulting sample lie in the range $1.4\le z\le 2.9$.

The multicolour photometry was obtained using the 4 bands HST+ACS data
in the optical (Giavalisco et al. 2004), the JHK bands in the near-IR
from VLT+ISAAC observations (Retzlaff et al. 2010) and from
Spitzer+IRAC (Dickinson et al., {\it in preparation}).  We used
PSF-matched images to build photometric catalogues from $B$ to $K$. The
IRAC photometry was measured over 4$''$ diameter apertures and
corrected to total magnitudes using corrections appropriate for point
sources, and matched to the $K$ band using total  $K$-band magnitudes as a
comparison. This is the same procedure that was used for the GOODS-S
catalogues presented by Cimatti et al. (2008) and Daddi et al. (2007a).
In summary, all SED fits make use of the following bands: $BVizJHK$ plus the Spitzer/IRAC channels 1, 2 and 3.

\begin{figure*}
\includegraphics[width=1.0\textwidth]{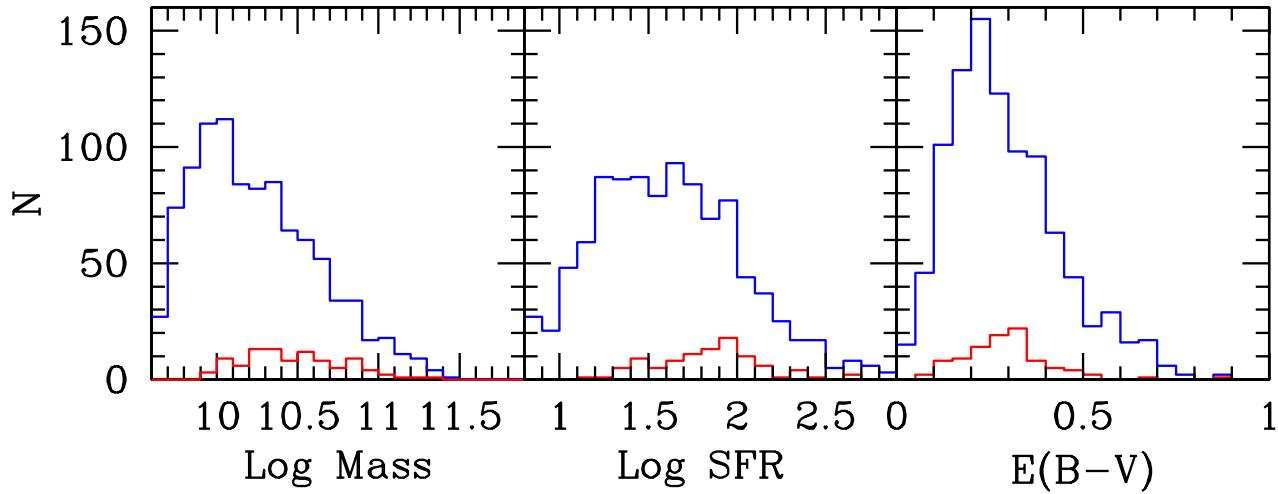}
\caption{The distribution functions of the stellar mass, SFR, and
reddening of $BzK$ galaxies in the GOODS-South sample as from Daddi et
al. (2007a), separately for the full sample, and for the sub-sample
with spectroscopic redshifts which is used in this paper.
\label{histo3}}
\end{figure*}

Besides satisfying the $BzK$ criterion, the sample of galaxies used in this paper is subject to the additional selection imposed by the various spectroscopic surveys mentioned above. Figure~\ref{histo3} shows the distribution functions of the mass, SFR, and reddening, as derived by Daddi et al. (2007a), separately for the full $K_{\rm Vega}<22$ sample in the GOODS-South field, and for the sub-sample of 96 objects with spectroscopic redshifts. The sample used here is somewhat biased towards higher masses, SFRs, and extinctions, but by and large for all three quantities it covers a major fraction of the range exhibited by the full sample. This can also be appreciated by inspecting Figure~\ref{sfrm}, where the 96 galaxies used in the present study are compared to the whole sample in Daddi et al. (2007a).  Therefore, we consider that the conclusions drawn from this spectroscopic sample should hold for the full sample, perhaps with the exclusion of some low mass, low SFR objects.

This database allows the sampling of galaxy SEDs up to the rest-frame
$K$ band. The SEDs over the whole wavelength range from the rest-frame
$UV$ to the $K$ band will be analysed in the next Sections.

\section{SED fitting}
\subsection{Generalities}

The method we adopt for the SED fitting is similar to the one used in
Maraston et al. (2006). We construct composite population templates
based on the stellar population models of Maraston~(2005) and we use
an adapted version of the code {\it Hyper-Z} (Bolzonella et al. 2000),
in which the SED fitting is performed at fixed spectroscopic
redshift\footnote{For the fitting procedure we use photometric errors
of 0.05 if the formal error is smaller than that, to account for
systematics in photometry and colour matching}. We use an updated
version of Hyper-$Z$, kindly provided to us by M. Bolzonella, in which
221 ages (or $\tau$'s) are used for each kind of SFH, instead of the
51 used in earlier versions. The use of denser grids tends to give
somewhat different results, an effect that is explored in detail in a
parallel paper (J. Pforr et al. in preparation).  

It is important to note that the code does not interpolate on the template grids, hence the template set must be densely populated.

The fitting procedure is based on maximum-likelihood algorithms and
the goodness of the fit is quantified via the $\chiquadr$
statistics. The code computes $\chiquadr$ for a large number of
templates, which differ for SFHs, and finds the best-fitting template
among them, having the reddening $E(B-V)$ as an additional free
parameter.  It is important to note that the code does not interpolate
on the template grids, hence the template set must be densely
populated.  The extinction $A_{\rm V}$ is allowed to vary from 0 to 3
in steps of 0.2, which corresponds to $E(B-V)$ from 0 to 0.74
according to the reddening law of Calzetti et al. (2000), that we
adopt for all fits. By doing so we implicitly assume that the
  dust composition is the same in all examined galaxies and that there
  are no major galaxy to galaxy differences in the relative
  distribution of dust and young, hot stars. Differences in
  the extinction curve have been detected among $z\sim 2$ galaxies,
  with some galaxies exhibiting the 2175 \AA\ UV bump, while others
  not showing it (Noll et al. 2009), but such differences appear
  to have only minor effects on the derived SFR. Extinction is
  unlikely to be uniform across the surface of galaxies, 
  particularly in extremely dusty ones (e.g., Serjeant, Gruppioni \& Oliver 2002). However, the
  tightness of the SFR-mass relation for $z\sim 2$ galaxies
  (cf. Fig. 1), together with the agreement of their UV-derived SFRs
  with those derived from the radio (Pannella et al. 2009) argues for
  such {\it average extinction} approximation to be a fairly good one,
  at least for the majority of the galaxies at these redshifts.

For all models we considered only solar metallicity. This is different from the approach adopted by Maraston et al. (2006) in fitting the near-passive galaxies at $z\lsim 2$, for which we considered four metallicities and several possible reddening laws.  Indeed, we have noticed that varying the reddening law has only a very mild influence on the derived properties of star-forning galaxies, hence for economy we decided to stick to the reddening law that gives the best-fit in most cases, which is the Calzetti law. As for the metallicity, it is known that super-solar metallicities tend to give younger ages and higher masses and vice-versa for sub-solar metallicities (e.g., Maraston 2005). However, metallicity effects do not change the main results of the present investigation. Indeed, our focus is on exploring the effects of adopting different functional forms for the SFHs and we restrict the main analysis at fixed, solar metallicity. On the other hand, metallicity effects over the SED of star forming galaxies are generally less important than age effects.

The main difference with respect to Maraston et al. (2006), which was
focused on nearly-passive galaxies, consists in the composite
population templates that are used in the fits. In particular, besides
a constant star formation and direct-$\tau$ models, we explore
inverted-$\tau$ models for various setups of ages and $\tau$'s. The
latter models are quite a novelty in this kind of studies, and we
describe them in more detail below.

\subsection{Direct and inverted $\tau$ models}

As mentioned above, besides constant SFR models, in this paper we consider two main sets of SFHs, namely direct-$\tau$ and inverted-$\tau$ models,
where the evolution of the SFR is given by Equation (2) and (3), respectively.
For both types of SFHs, we calculate three model
flavours which mostly differ with respect to how the parameter age is
treated, namely:
\par\noindent
a) leaving the age as a free parameter; 
\par\noindent
b) fixing the age by fixing the formation redshift and varying only $\tau$;
\par\noindent
c) constraining the age to be larger than $\tau$.
\par\noindent
The age of composite models is the time elapsed since the
beginning of star formation, i.e. is the age of the oldest stars
$(t-t_\circ)$.  In the age free case, we simply compute exponentially
increasing/decreasing models for various $\tau$'s, and 221 ages for each
$\tau$. The SED fit with these age-free models releases for each galaxy a value
of age $=t-t_\circ$, $E(B-V)$, $M_*$ and of $\tau$.

In the case of fixed age (case b), we assume that all galaxies started
to form stars at the same redshift $z_{\rm f}$, and therefore the
age of a galaxy follows from the cosmic time difference between its
individual spectroscopic redshift and $z_{\rm f}$. Thus, in this
case the best fit releases a value of $\tau$, $E(B-V)$ and $M_*$.  In
this experiment we take $z_{\rm f} = 5$ for the formation redshift, which implies that all
galaxies are older than $\sim ~ $ 1 Gyr (ages range between 1 and 3
Gyr, depending on their redshift).

In case c) the age can still vary freely, but only within values that
exceed the corresponding $\tau$, so e.g. when $\tau$~is 0.5 Gyr, the
allowed ages for the fit must be $\ge 0.5$ Gyr. In practice, among all
the fits attempted for case a) those with age $<\tau$ are not considered.

Age is a default free parameter in Hyper-$Z$, and therefore the
procedure had to be modified to cope with case b), for which age is no
longer a free parameter.  Thus, for each age we have calculated a grid
of models for 221 different $\tau$'s ranging from 0.05 to 10.3 Gyr in
step of 0.045 Gyr, although the cases with $\tau<0.3$ Gyr will be discussed separately.
The best-fit then finds the preferred
$\tau$. If we were to consider all kinds of galaxies,
including passive galaxies and low-redshift ones, the
exponentially-increasing models should be quenched at some point,
which could be done either by SFR truncation, or adding a further
exponential decline. We did not quench the models here as the galaxies
we focus on are all actively starforming at the epoch of
observation. It is worth emphasising that the set-up with fixed age clearly
has one degree of freedom less than those with age free, which will affect
their $\chiquadr$ values.

The quality of the fits is measured, as usual, by their reduced
$\chi^2$, but we believe that the plausibility of the fits can be
assessed only considering the broadest possible astrophysical
context. Before presenting our results it is worth adding some final
comments on the derived stellar masses and SFRs.  In this work we
have used templates adopting a straight Salpeter IMF down to 0.1
$M_{\odot}$. This is not the optimal choice as it is generally agreed
that an IMF with a flatter slope or cut-off at low masses, like those
proposed by Kroupa (2001) or Chabrier (2003), is more appropriate for
deriving an absolute value of the stellar mass or of the SFR.
However, the focus of this work is in comparing the results obtained
with different SFHs, hence the slope of the IMF below $\sim 1\,\msun$
is irrelevant. Note also that the reported stellar mass $M_{*}$ is the
mass that went into stars by the age of the galaxy. This overestimates
the true stellar mass, as stars die leaving remnants whose mass is
smaller than the initial one, and the galaxy mass decrement in case of
extended star formation histories is $\sim 20-30\%$ (see e.g.,
Maraston 1998; Maraston et al. 2006).
\section{Results}
In this Section we compare the stellar population properties derived
under the three different assumptions for the SFH, namely: direct- and
inverted-$\tau$ models and constant star formation, both leaving age
as a free parameter,  and also assuming age (i.e., formation redshift) as a
prior, as described above.
\subsection{Age as a free parameter}
A first set of best fits was performed allowing the procedure to
select the preferred galaxy age. The results are reported in
Figures 3 to 8, showing the histograms for the resulting ages,
$\tau$'s, stellar masses, reddening, SFRs and reduced $\chi^2$,
respectively.
\begin{figure}
\includegraphics[width=0.49\textwidth]{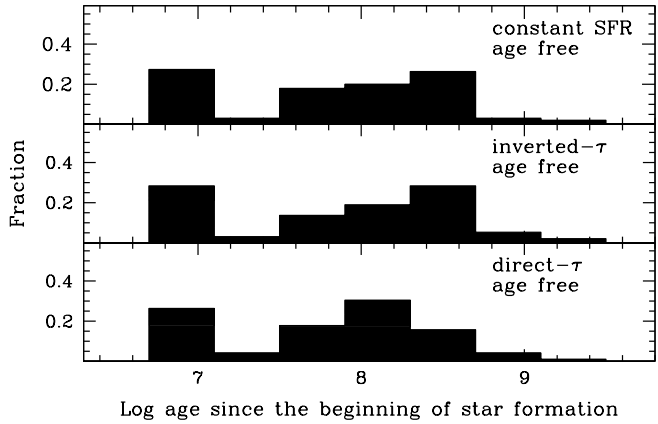}
\caption{The time elapsed since the beginning of star formation, for
models with constant star formation (upper panel), and with star
formation proceeding as an inverted and direct-$\tau$ model (middle
and lower panels, respectively). The age is left free in all three models. \label{ages}}
\end{figure}
\begin{figure}
\includegraphics[width=0.49\textwidth]{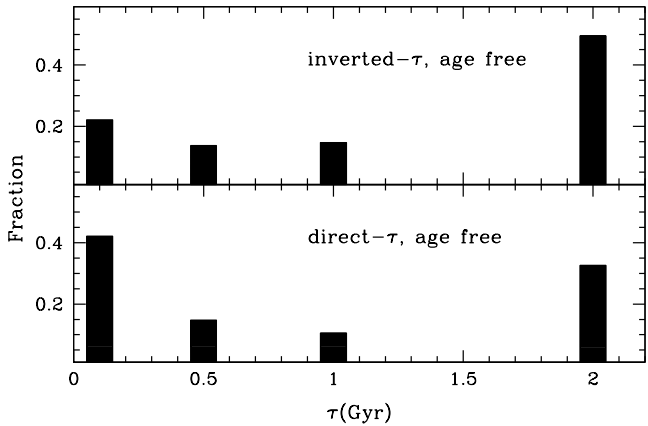}
\caption{Same as Figure 3 for $\tau$ (in Gyr). \label{taus}}
\end{figure}
\begin{figure}
\includegraphics[width=0.49\textwidth]{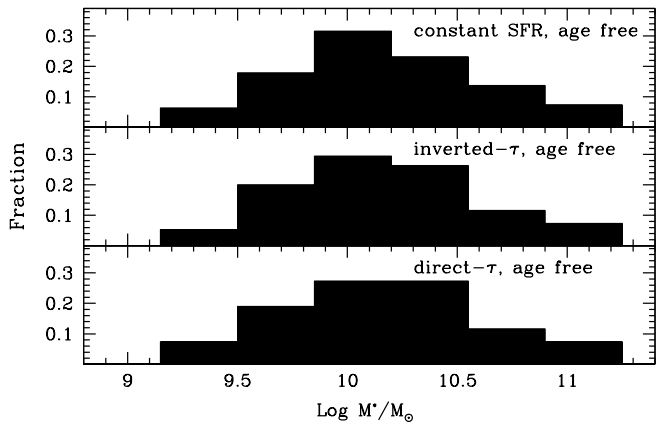}
\caption{Same as Figure 3 for stellar masses. \label{mass}}
\end{figure}
\begin{figure}
\includegraphics[width=0.49\textwidth]{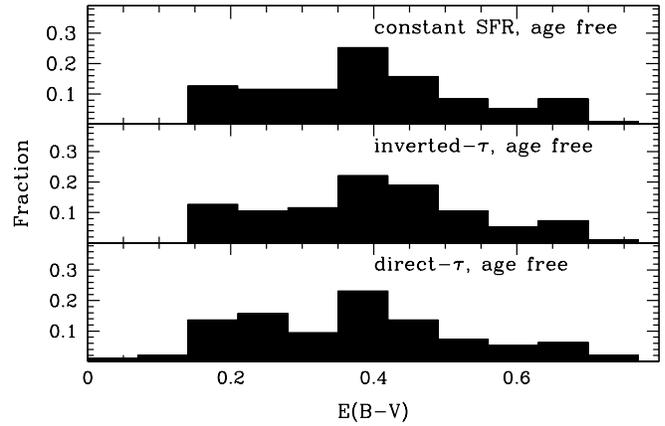}
\caption{Same as Figure 3 for the reddening $E(B-V)$. \label{ebv}}
\end{figure}

\begin{figure}
\includegraphics[width=0.49\textwidth]{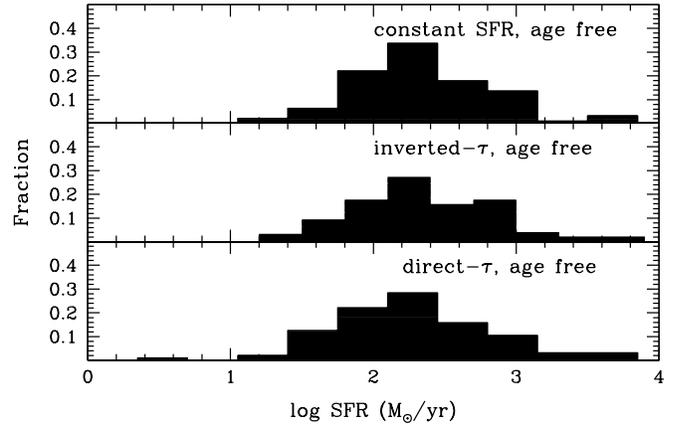}
\caption{Same as Figure 3 for the star formation rates. \label{sfrhisto}}
\end{figure}
\begin{figure}
\includegraphics[width=0.49\textwidth]{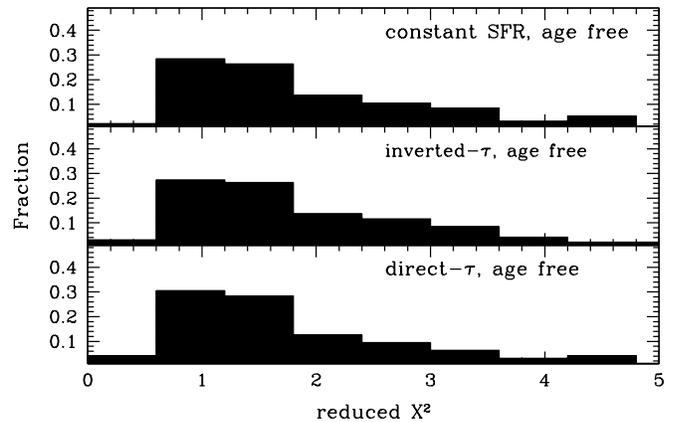}
\caption{Same as Figure 3 for the \chiquadr. 
\label{chihisto}}
\end{figure}
\begin{figure}
\includegraphics[width=0.49\textwidth]{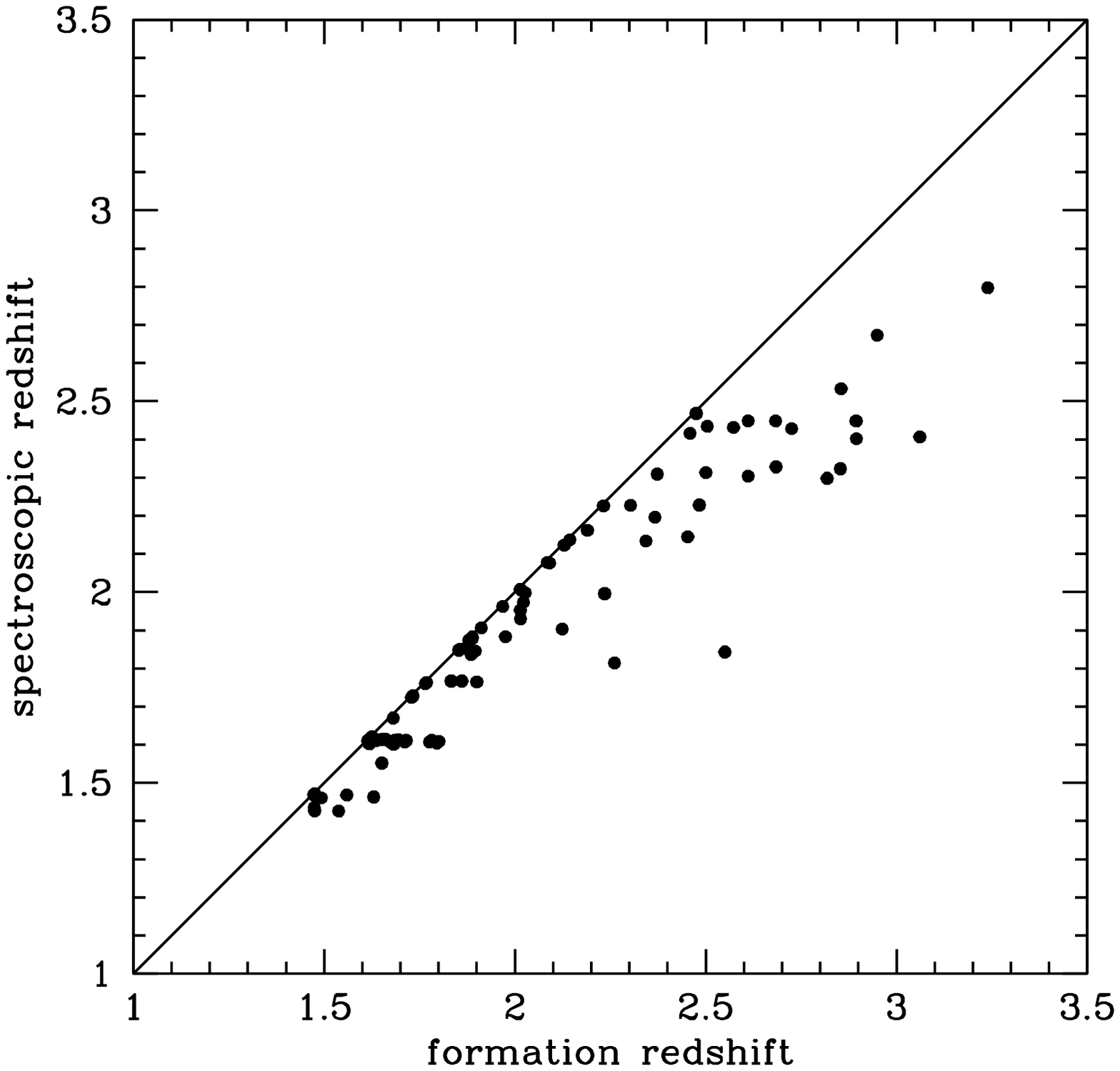}
\caption{The spectroscopic redshift of individual galaxies vs, their
{\it formation redshift} as deduced from their age as derived from
inverted-$\tau$ models with age as a free parameter. Similar
plots could be shown using ages derived from direct-$\tau$ models or
models with SFR=const. \label{zzform}}
\end{figure}

Surprisingly, all histograms look quite similar, irrespective of the adopted SFH. In particular, the distributions of the reduced $\chi^2$ values are almost identical, indicating that no adopted functional shape of the SFH gives substantially better fits than another. However, it would be premature to conclude that the resulting physical quantities (ages, masses, etc.) have been reliably determined. Most of the derived ages are indeed very short in all three cases, with the majority of them being less than $\sim 2\times 10^8$ yr, with several galaxies appearing as young as just $\sim 10^7$ yr (see Figure 3). We believe that these ages - intended to indicate the epoch at which the star formation started - cannot be trusted as such, for the following reasons. First, as we shall show later and as also known in the literature, the latest episode of star formation outshines older stars even when composite models are considered. In addition, cosmological arguments render this interpretation suspicious, as we are going to discuss. Figure~\ref{zzform} shows the redshift of each individual galaxy as a function of its formation redshift, the latter being derived by combing the redshift and age of each galaxy. Note that with the derived ages most galaxies would have started to form stars just shortly before we happen to observe them. If this were true, the implied cosmic SFR should rapidly vanish by $z\sim 3$, while this is far from happening: the cosmic SFR at $z\sim 3$ is nearly as high as it is at $z\sim 2$ (e.g., Hopkins \& Beacom 2006), and the specific SFR at a given mass stays at the same high level all the way to $z\sim 7$ (Gonzalez et al. 2010). 

We consider far more likely that most of the descendants of galaxies responsible for the bulk of cosmic star formation at $z\gsim 3$ are to be found among the still most actively star forming galaxies at $z\sim 2$, rather than a scenario in which the former galaxies would have faded to unobservability by $z\sim 2$, and those we see at $z\sim 2$ would have no counterpart at $z\gsim 3$.  In other words, massive star forming galaxies at $z\sim 2$ must have started to form stars at redshifts well beyond $\sim 3$.  Note that some star forming galaxy at $z\gsim 3$ may have turned passive by $z\sim 2$, but the number and mass density of passive galaxies at $z\sim 2$ is much lower than that of $z\gsim 3$ star forming galaxies of similar mass (e.g. Kong et al. 2006; Fontana et al. 2009; Williams et al. 2009; Wuyts et al. 2009b).

\begin{figure*}
\includegraphics[width=0.82\textwidth]{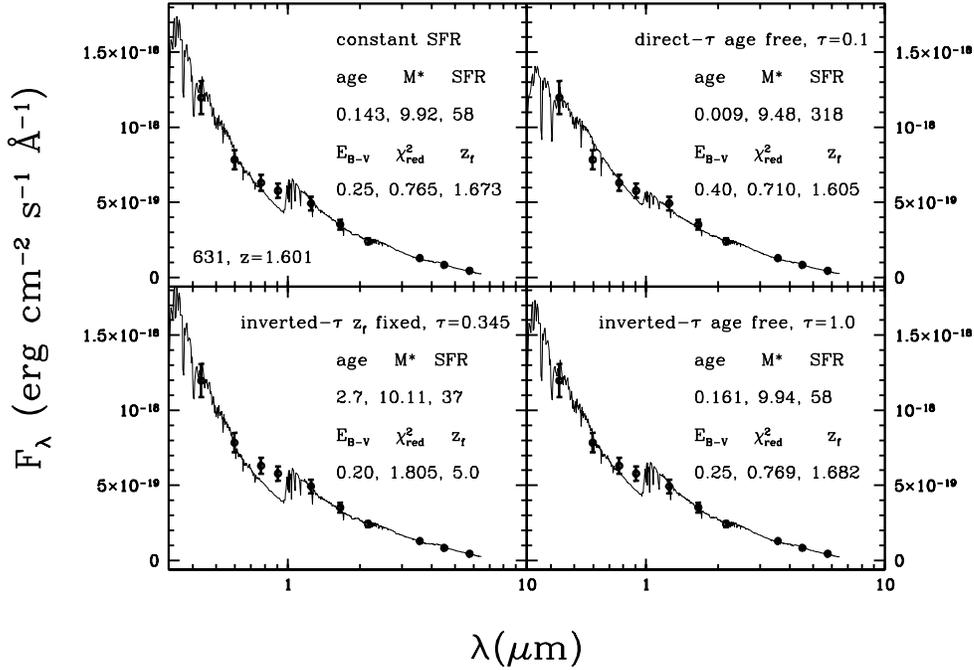}
\caption{The observed spectral energy distributions (filled symbols
with error bars) of high-$z$ starforming galaxies. Object ID and
spectroscopic redshift are labelled in the top left-hand panel. In
each panel the red line corresponds to the best fit solution according
to the different templates for the star formation history, namely
constant star formation, $\tau$ with age free, inverted-$\tau$ with
age fixed and inverted-$\tau$ with age free (from top to bottom and
left to right). Several parameters of the fits are labelled, namely
the age (in Gyr), the $\tau$ (in Gyr), the reddening $E(B-V)$, the
stellar mass (in log $M_\odot$), the Star Formation Rate (SFR) in
$M_\odot$/yr, the \chiquadr, the 'formation redshift' $z_{form}$ obtained
by subtracting the formal age of the object to the look-back time at
given spectroscopic redshift. \label{sedfit}}
\end{figure*}

\begin{figure*}
\includegraphics[width=0.82\textwidth]{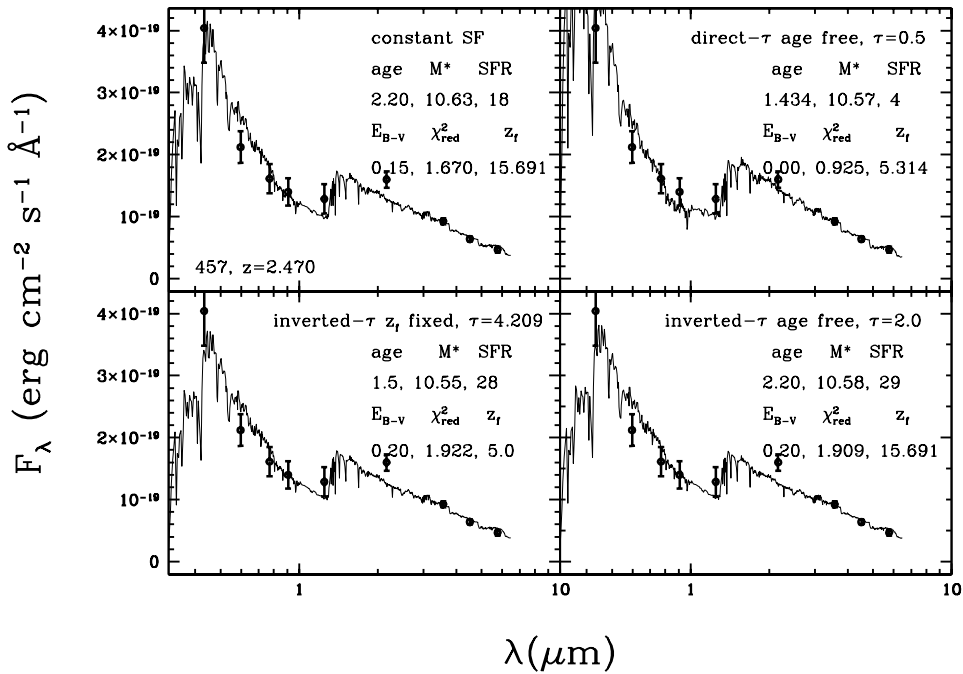}
\caption{The same as in Figure~\ref{sedfit}, for one of the few
galaxies with best fit age $\sim 1$ Gyr. Note the sizeable Balmer
break. \label{sedfit457}}
\end{figure*}

Thus, the question is why does the best fit procedure choose
such short ages?  Clearly, the result is not completely unrealistic,
as the bulk of the light must indeed come from very young stars. This
is illustrated by the example shown in Figure~\ref{sedfit}, for a
typical case in which very short ages are returned. The main
age-sensitive feature in the spectrum is the Balmer break, which for
this galaxy as for most others in the sample, is rather weak or absent. With
the whole optical-to-near-IR SED being dominated by the stars having formed in the recent
past, the spectrum does not convey much age information at
all. Figure~\ref{sedfit}, lower/left panel, shows that if one forces
age to be as large as 1.5 Gyr (and fix the formation redshift
accordingly), then the Balmer break deepens, and the $\chiquadr$
worsens. Figure~\ref{sedfit457} shows the SED and relative best fit
spectra for one of the few galaxies for which the procedure indicates
a large age ($\sim 1$ Gyr). Clearly, this larger age follows from the
much stronger Balmer break present in this galaxy.

Note that for both direct- and inverted-$\tau$ models, in most cases
$\tau >> (t-t_\circ)$, i.e., the e-folding time of SFR is (much)
longer than the age, i.e., the SFR does not change much within the
time interval $t-t_\circ$. This explains why both kind of models give
results so similar to those in which SFR is assumed to be
constant. 

We shall show in the next subsections that other fits may actually
result in more plausible physical solutions, even if they have a {\it
worse} \chiquadr.

\begin{figure*}
\includegraphics[width=1.\textwidth]{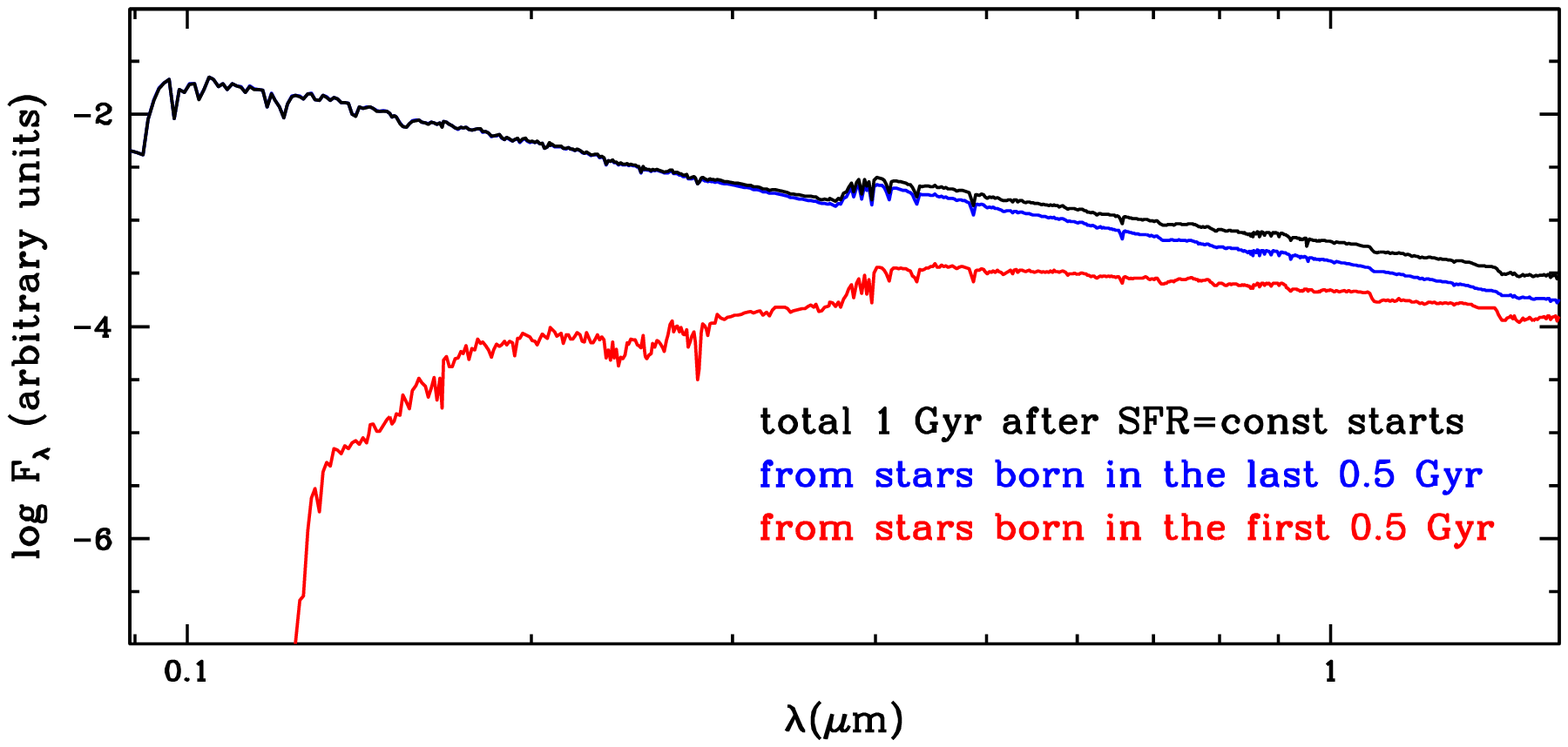}
\caption{The effect of outshining by the youngest fraction of a
composite population: the synthetic spectrum is shown of a
composite population having formed stars at constant rate for 1
Gyr. The contribution of the stars formed during the first and the
second half of this period are shown separately as indicated by the
colour code, together with the spectrum of the full population.
\label{overshine}}
\end{figure*}

The extremely young ages and
the insensitivity of the fit to different star formation histories
when the age is left free should not be so surprising. They are
the consequence of the fact that the very young, massive stars just
formed during the last $\lsim$ few $10^8$ yr dominate the light at
virtually all wavelengths in these actively star forming galaxies,
even if they represent a relatively small fraction of the stellar mass.  The
capability of a very young population to outshine previous stellar
generations is illustrated in Figure~\ref{overshine}. Synthetic
spectra are shown for a composite stellar population that has formed
stars at constant rate for 1 Gyr, and then separately for the contributions
of the stars formed during the first half and the second half of this
time interval. The contribution of the young component clearly outshines the
old component at all wavelengths, making it difficult to assess the
presence and contribution of the latter one, even if it represents
half of the total stellar mass.  This plot demonstrates why it is not
appropriate to interpret the ``age'' resulting from the previous fits
as the time elapsed since the beginning of star formation, as it is
formally meant to be in the fitting procedure and as might be suggested by a
naive interpretation of the fitting results. It would instead be better interpreted as the age of the stars producing
the bulk of the light. In the next
subsection we experiment with a different approach that may circumvent this
intrinsic problem and get more robust properties of (high-redshift)
star-forming galaxies.
\subsection{Best fit solutions with fixed formation redshift}
Clearly massive star-forming galaxies at $z\sim 2$ cannot have started
to form stars just shortly before we observe them. Instead,
they must have started long before, as indeed demonstrated by the fact
that starforming galaxies are found to much higher
redshifts. In this section we arbitrarily assume that all our galaxies
have started to form stars at the same cosmic epoch, corresponding to
$z=5$, i.e. when the Universe was $\sim 1$ Gyr old. This
implies an age for the galaxy at redshift $z$ which is given by
$\Delta t = t(z)-t(z_{\rm f})$. At first sight such an assumption may
seem a very strong one. Actually this is not the case. We have already
argued that our galaxies must have started to form stars at a redshift
much higher than that at which they are observed, i.e., $z_{\rm
f}>>2$. Therefore, the prior age changes only marginally (few percent)
if instead of $z_{\rm f}=5$ we had chosen $z_{\rm f}=6$, 7, or more,
and so does the result of the fit, i.e., such result is fairly
insensitive to the precise value of $z_{\rm f}$, provided it is well
in excess of $\sim 2$. Obviously, the effect is especially strong for exponentially increasing SFRs. Incidentally, we note that exponentially decreasing SFRs
have been assumed also for galaxies at very high redshifts (e.g., $z=4-6$;
Stark et al. 2009).

\begin{figure}
\includegraphics[width=0.40\textwidth]{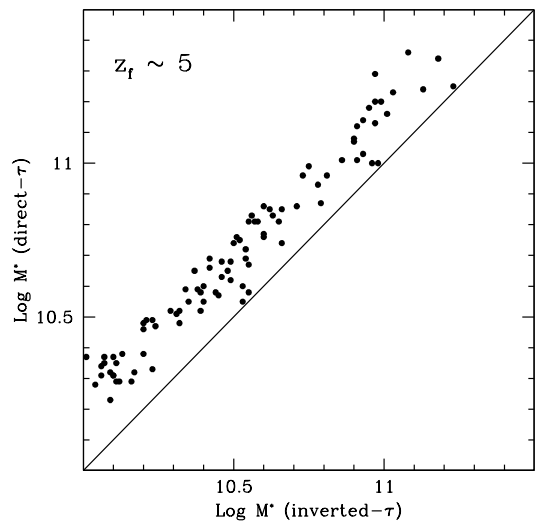}
\caption{A comparison of the stellar masses derived assuming
exponentially decreasing star formation rates (direct-$\tau$ models)
with those obtained assuming exponentially increasing star formation
rates (inverted-$\tau$ models).  The beginning of star formation is set
at the cosmic epoch corresponding to $z=5$ for all models.
\label{masszf5}}
\end{figure}

\begin{figure}
\includegraphics[width=0.40\textwidth]{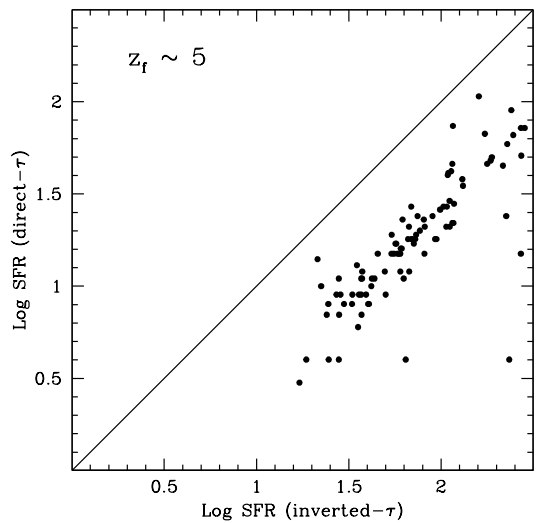}
\caption{The same as in Figure~\ref{masszf5}, but for a comparison of
the derived star formation rates.
\label{sfrzf5}}
\end{figure}

\begin{figure}
\includegraphics[width=0.40\textwidth]{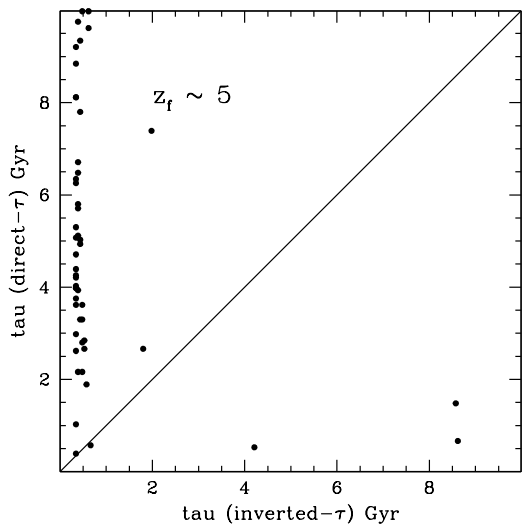}
\caption{The same as in Figure~\ref{masszf5}, but for a comparison of
the derived $\tau$ values.
\label{tauzf5}}
\end{figure}

\begin{figure}
\includegraphics[width=0.40\textwidth]{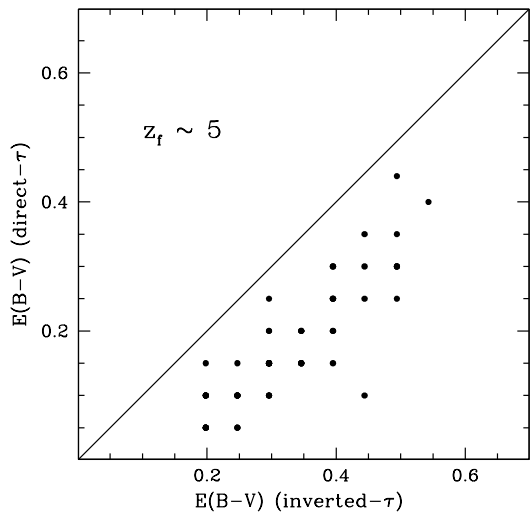}
\caption{The same as in Figure~\ref{masszf5}, but for a comparison of
the derived values of the reddening $E(B-V)$ .\label{ebvzf5}}
\end{figure}

For both the exponentially declining and increasing $\tau$ models, we identify the best-fitting SFHs, first allowing only values of $\tau\ge 0.3$ Gyr.  The results are shown in Figures~\ref{masszf5}-\ref{ebvzf5}, in which the parameters of the best-fitting models for these two SFHs are compared.

\begin{figure}
 \includegraphics[width=0.40\textwidth]{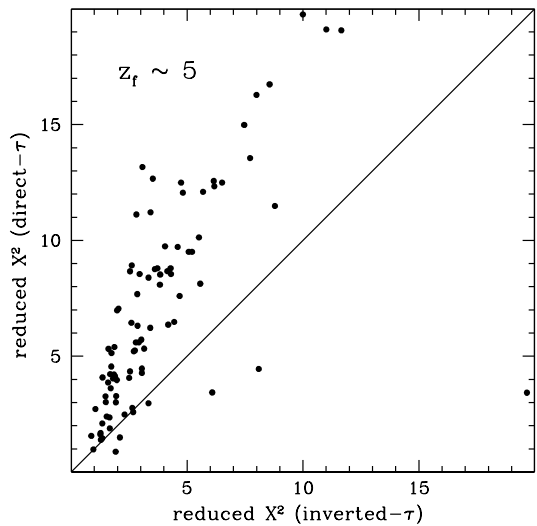}
\caption{The same as in Figure~\ref{masszf5}, but for a comparison of
the reduced $\chi^2$ values of the best fit solutions.
\label{chizf5}}
\end{figure}

Figure~\ref{masszf5} compares the stellar masses obtained with the two
SFHs, showing that those obtained with direct-$\tau$ models are systematically
larger by $\sim 0.2$ dex with respect to stellar masses derived using the
inverted-$\tau$ models. Figure~\ref{sfrzf5} shows the comparison of
the SFRs, with the SFR from inverted-$\tau$
models being systematically higher by $\sim 0.5$ dex, and therefore
the specific SFR turns out to be systematically higher by $\sim
0.7$ dex.  Figure~\ref{tauzf5} shows how different the $\tau$
values derived using the two SFHs are. Direct-$\tau$ models prefer very
large $\tau$'s, up to $\sim 10$ Gyr, which is to say that they prefer
nearly constant SFRs. On the contrary, inverted-$\tau$ models prefer
short $\tau$'s, typically shorter than $\sim 0.5$ Gyr, hence a SFR
that is rapidly increasing with time.

Figure~\ref{tauzf5} gives the key to understand the physical origin of
the mass and SFR offsets between the two families of models. Being
forced to put more mass at early times, direct-$\tau$ models pick very
large $\tau$'s trying to find the best compromise for mass and SFR,
and, compared to inverted-$\tau$ models,  they overproduce mass at early times and not enough star
formation at late times. On the other hand, these
latter models by construction put very little
mass at early times, and most of it is formed at late times. In order
to compromise mass and SFR, they may overestimate the current SFR, and
may be forced to hide part of it demanding more extinction.

Indeed, Figure~\ref{ebvzf5} compares the reddening obtained with the two
opposite SFH assumptions, showing that to obtain a good fit in the
rest-frame UV the higher ongoing SFR derived from inverted-$\tau$ models needs
to be more dust-obscured than in the case of direct-$\tau$
models. Finally, Figure~\ref{chizf5} compares the reduced $\chi^2$'s
obtained with the two sets of models. Those relative to
inverted-$\tau$ models are not very good, but those with direct-$\tau$
models are definitely much worse. Still, a robust choice for the SFH
cannot rely only on this relatively marginal advantage of
inverted-$\tau$ models. In the next subsections we try to gather
independent evidence that may help favouring one or the other option.

\subsection{Comparing SFRs and extinctions from SED fitting and from 
rest-frame UV only}
The rest-frame UV part of the explored spectrum (from observed
$B$ band up to Spitzer/IRAC channel 3, i.e., 5.8 $\mu$m) is the one
which most directly depends on the rate of ongoing star formation. On
the other hand, when the SFR is estimated with an SED-fitting
procedure as in the previous sections, the derived SFR results from
the best-possible compromise with all the free parameters, given the
adopted templates. In other words, the resulting SFR is {\it
compromised} relative to the other free parameters of the fit and
the adopted SFHs. SFRs from the UV flux, corrected for extinction
using the UV slope (plus a reddening law, such as the Calzetti law,
Calzetti et al. 2000) are widely derived in the literature for
high redshift galaxies, and shown to be in very good agreement with
independent estimates from the radio flux at 1.4 GHz (e.g., Reddy et
al. 2004; Daddi et al. 2007a,b; Pannella et al. 2009), and also from
the mid-IR (24 $\mu$m) and soft X-ray fluxes, for those galaxies with
no mid-IR excess (Daddi et al. 2007b). Thus, the SFR derived only from
the UV flux represents an {\it uncompromised} template (in the
sense that it has no covariance with other parameters of the fit), and
therefore with respect to which we may gauge the physical plausibility of a {\it whole optical-near-IR SED} fit.

\begin{figure*}
\includegraphics[width=0.7\textwidth]{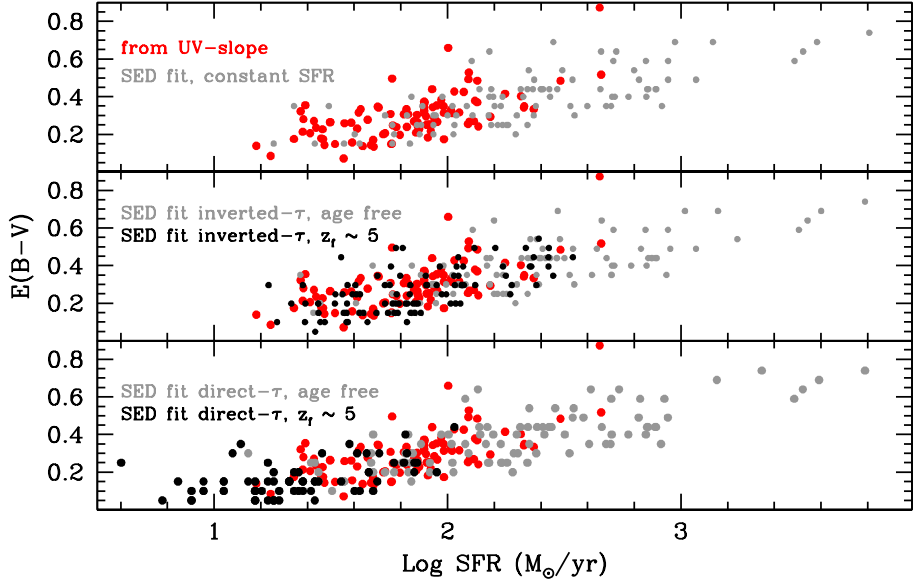}
\caption{The reddening and SFRs derived under
different assumptions on the SFH are compared to those derived using
only the rest-frame UV part of the spectrum (red points), i.e., the
wavelength range that most directly depends on the ongoing rate of
star formation. The upper panel shows a comparison of such UV-derived
SFRs with those derived from SED fitting assuming SFR=const and
leaving age as a free parameter. In the middle panel the comparison is
made with inverted-$\tau$ models, with both fixed and free age (black
and grey points, respectively). Finally, the lower panel shows the
comparison with the direct-$\tau$-models.
\label{sfrebv}}
\end{figure*}

Figure~\ref{sfrebv} compares the $E(B-V)$ vs. SFR relations that are
obtained from SED fittings with various SFHs, to those obtained using
only the $UV$ part of the spectrum. The latter ones were derived by
Daddi et al. (2004, 2007a) using the Bruzual \& Charlot (2003) models for
converting $L_{\rm UV}$ into SFR, following Madau et al
(1998). $E(B-V)$ is derived by mapping the $(B-z)$ colour into
$E(B-V)$ using the Calzetti et al. law. Note that there are no
appreciable differences in the UV between the models of Bruzual \&
Charlot (2003) and those of Maraston (2005) that are used in this
paper. As clearly shown, the best agreement is between UV-derived SFRs
and those obtained from SED fitting with inverted-$\tau$ models and
fixed formation redshift.  Notice that when leaving age as a free
parameter several galaxies are found with exceedingly large SFRs and
$E(B-V)$ values (see also Section 4.1), much at variance with the more
robust values derived from the UV flux. Direct-$\tau$ models with
fixed formation redshift are in better agreement with the UV-derived
SFRs, yet they systematically underestimate the SFR as already noticed
in Section 4.2.

\begin{figure*}
\includegraphics[width=0.9\textwidth]{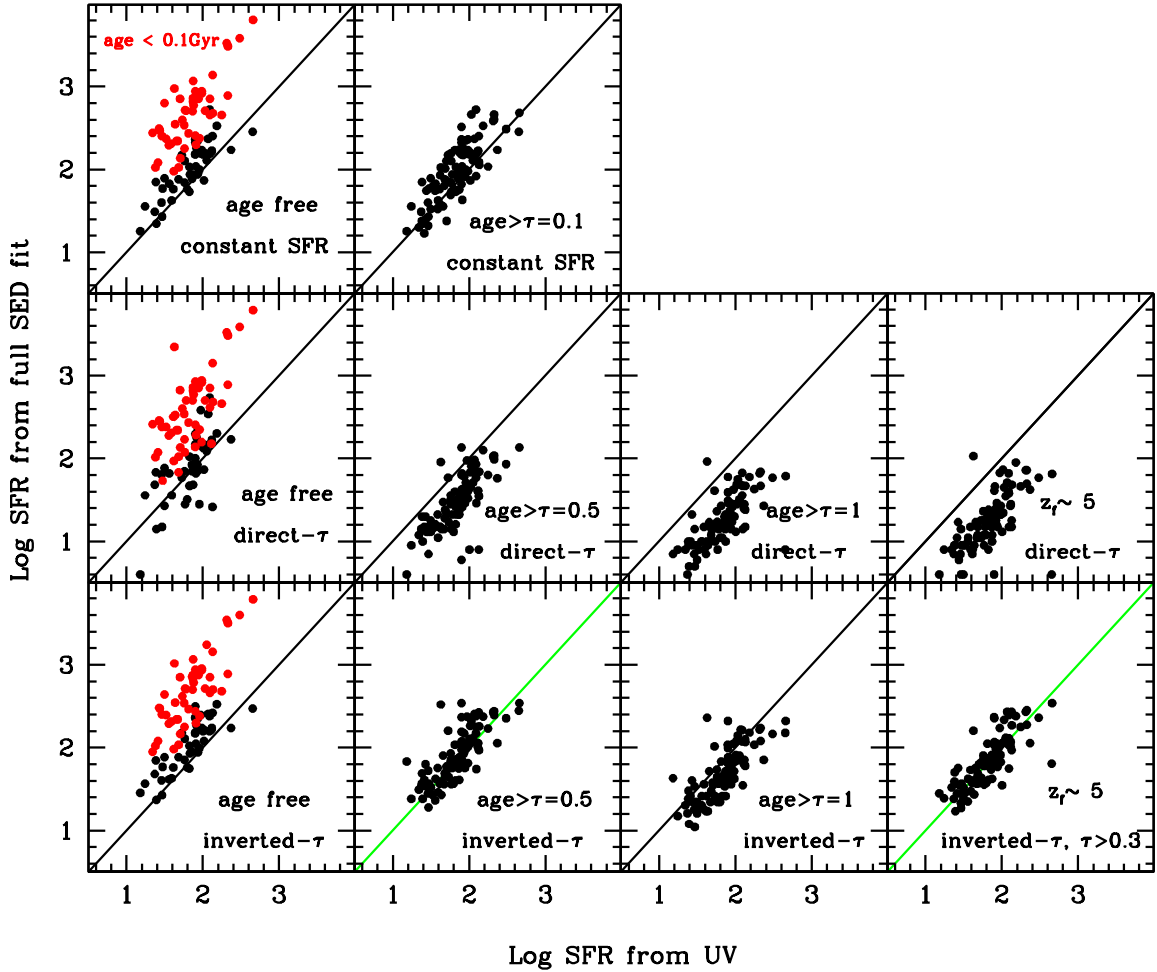}
\caption{The comparison of SFRs derived from SED fitting with those
derived from the rest-frame UV only (plus extinction correction).
Inverted-$\tau$ models with $z_{\rm form}=5$ (or inverted-$\tau$s with
fixed $\tau$=0.5 Gyr and ages constrained to be larger than $\tau$)
provide the best result (panels with green lines).
\label{sfrcomp}}
\end{figure*}
\begin{figure*}
\includegraphics[width=0.9\textwidth]{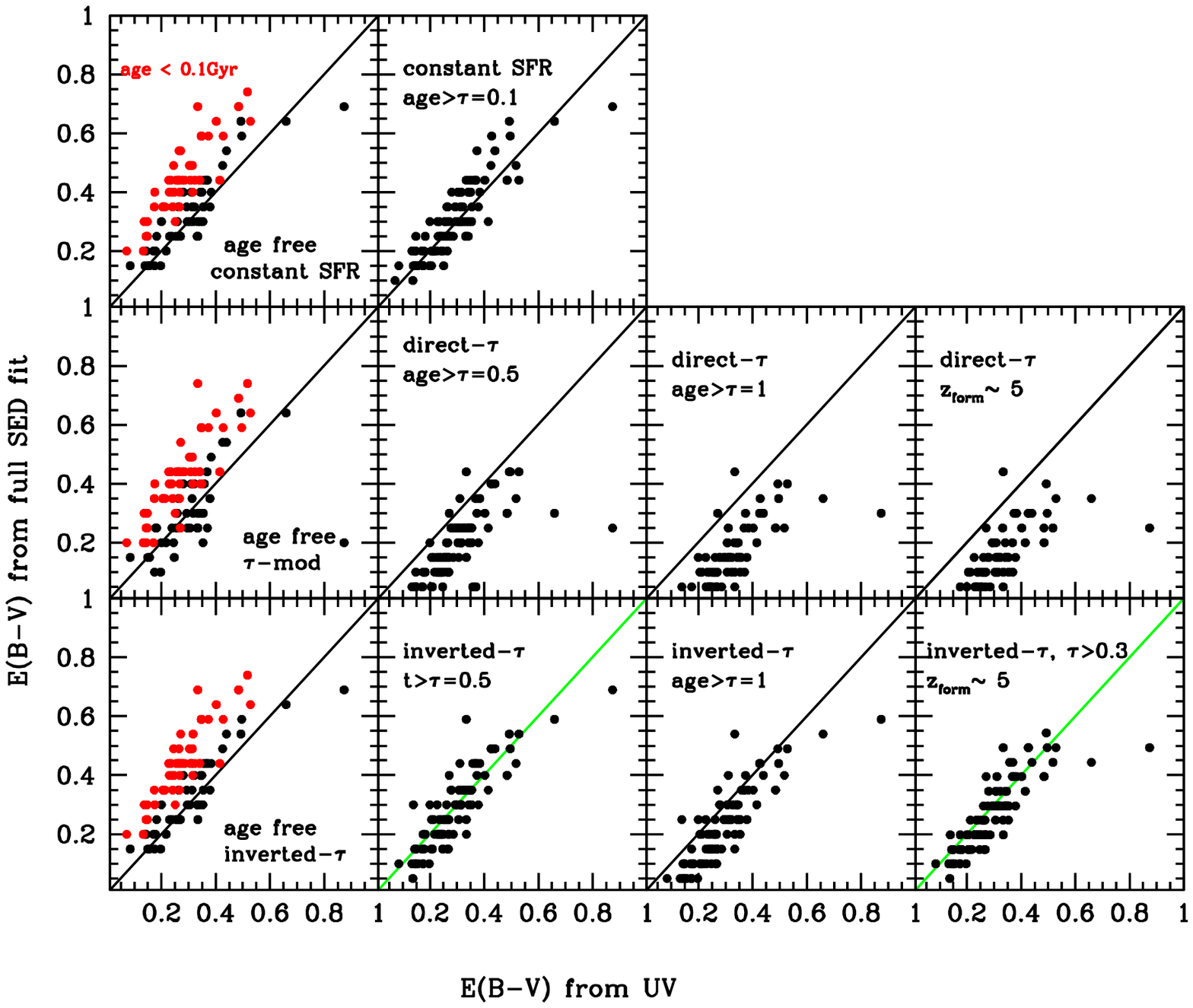}
\caption{The same as Figure~\ref{sfrcomp} for the reddening $E(B-V)$. Conclusions are
identical.\label{ebvcomp}}
\end{figure*}

This is further illustrated in Figure~\ref{sfrcomp} and
Figure~\ref{ebvcomp} which show the SFR and the reddening $E(B-V)$ from
the SED fits directly compared to those from the UV. Among all explored SFHs,
the inverted-$\tau$ models clearly are in best agreement with the values 
derived from the UV flux and slope. We interpret this as an indication 
that the SFHs of inverted-$\tau$ models models are closer to those of real 
galaxies, compared to the SFHs of direct-$\tau$ models.

\subsection{Best fit solutions with fixed $\tau$ and age $>\tau$}

It is easy to realise that in galaxies that follow a SFR such as that
given by Equation (1) the SFR tends to increase exponentially with
time (Pannella et al. 2009; Renzini 2009). More precisely, if one
ignores the $t$ term in this equation the exponential increase
proceeds with a $\tau \simeq 0.7$ Gyr. Thus, in this Section we present the
results of performing a new set of best fits, this time assuming a
fixed $\tau$ (namely $\tau$=0.5 and 1 Gyr, that bracket the empirical
value), leaving age as a free parameter but constraining it to be
larger than $\tau$.  We also consider the case of direct-$\tau$ models
with the same parameters and restrictions. This choice to set limits on
age may appear rather artificial, and indeed it is purely meant to
avoid the extremely small ages that are found when age is left
completely free. In practice, this exercise consists in considering
only the cases with age $>\tau$ among those already explored in
Section 4.1.

\begin{figure}
\includegraphics[width=0.49\textwidth]{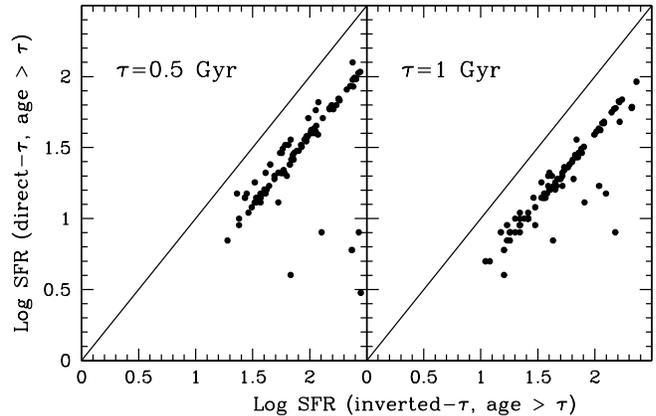}
\caption{Comparison of the star formation rates derived assuming
exponentially decreasing star formation rates (direct-$\tau$ models)
with those obtained assuming exponentially increasing star formation
rates (inverted-$\tau$ models). In both models, $\tau$~has been fixed
to two values of 0.5 and 1 Gyr, and the age is constrained to be
larger than $\tau$.
\label{sfr_agelargertau}}
\end{figure}

The comparison of the SFRs derived from direct- and inverted-$\tau$
models (Figure~\ref{sfr_agelargertau}) is qualitatively similar to the
case of fixed formation redshift (cf. Figure~\ref{sfrzf5}), and the
same comments apply also here.  The SFR and $E(B-V)$ values that are
obtained with these further models are then compared to those obtained
from the UV in Figures~\ref{sfrcomp} and \ref{ebvcomp}. It is apparent
that inverted-$\tau$ models with $\tau$=0.5 Gyr and ages larger than
$\tau$ give results nearly as good as those derived for the case of
fixed formation redshift. Also the case of
SFR = constant and age $>$ 0.1 Gyr SFRs results in reasonable agreement
with those derived from the UV.
All other explored SFHs give results at
variance from those obtained from the UV, that we regard as the most
robust method to estimate the SFR and reddening.

\begin{figure}
\includegraphics[width=0.49\textwidth]{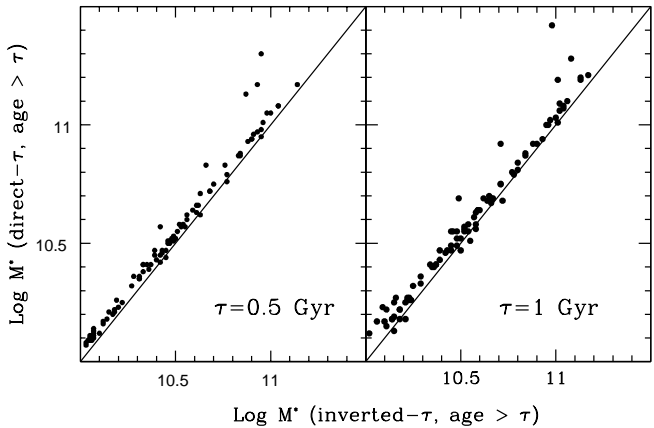}
\caption{The same as in Figure~\ref{sfr_agelargertau}, but for a
comparison of the derived stellar masses.
\label{mass_agelargertau}}
\end{figure}

\begin{figure}
\includegraphics[width=0.49\textwidth]{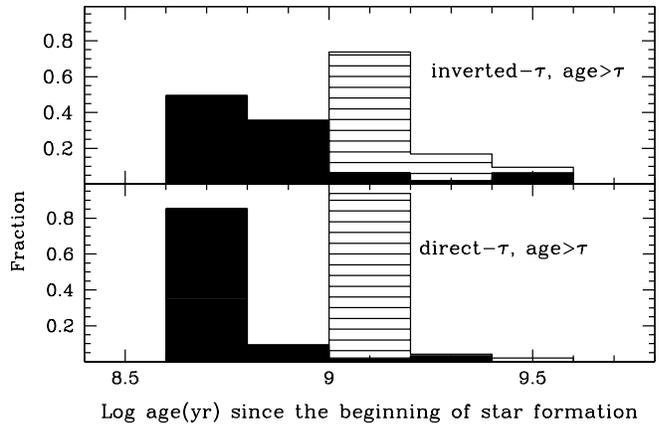}
\caption{Distributions of the ages obtained with models with fixed
$\tau$ and ages constrained to be larger than $\tau$, for $\tau=0.5$
Gyr (black histogram) and $\tau-1$ Gyr (shaded
histogram). 
\label{age_agelargertau}}
\end{figure}

Figure~\ref{mass_agelargertau} shows the derived stellar masses. In
this parameterisation there is less difference between the masses that
are derived with the age $>\tau$ constraint with respect to the case
of fixing the formation redshift. This is due to the fact that
direct-$\tau$ models indicate lower masses compared to those in the
case of fixed formation redshift, because of their lower age, hence
shorter duration of the star formation activity.  Note, however, that
in most cases the smallest $\chiquadr$ are obtained for the smallest
possible age, i.e., age$\simeq\tau$, or $\sim 0.5$ Gyr, as shown in
Figure~\ref{age_agelargertau}, again a consequence of the outshining effect.

\subsection{Blind diagnostics of the star formation history of mock galaxies}

The ability of an adopted shape of the SFH to recover the basic
properties of a composite stellar population can be tested on mock
galaxies with known SFHs.  In a parallel project (J. Pforr et al.,
{\it in preparation}) we use synthetic galaxies from semi-analytic
models (GALICS, Hatton et al. 2003) in which the input SSPs are the
Maraston (2005) models in the rendition of Tonini et al. (2009, 2010).
Their observed-frame magnitudes at the various redshifts are then
calculated, fed into Hyper-$Z$ just as if they were relative to real
observed galaxies, and best fits are sought using various template
composite stellar populations.\footnote{A similar project is described
in Wuyts et al. (2009a), and a detailed comparison with their results
is given in Pforr et al., in preparation. The systematic offsets
introduced by the use of direct-$\tau$ models to derive masses and
SFRs of mock galaxies at $z>3$ are also investigated by Lee et
al. 2009).}  Here we focus on an experiment involving the templates
used in this paper, namely inverted- and direct-$\tau$ models and
constant SF models, applied to mock star-forming galaxies at redshift
2. The model spectra are reddened according to the ongoing SFR, using
the empirical evidence that the $E(B-V)$ is proportional to the SFR
(e.g., Daddi et al. 2007a) of each individual galaxy, with
$E(B-V)=0.33\times ({\rm log\; SFR}-2)+1/3$. Extinction as a
function of wavelength is then applied following the law of Calzetti
et al. (2000). Note that we do not take such mock galaxies as
representative of real galaxies. The goal of the experiment is to show
what happens if the SED of a galaxy with a certain SFH is used to
derive its SFR and stellar mass using a different SFH.

\begin{figure*}
\includegraphics[width=0.9\textwidth]{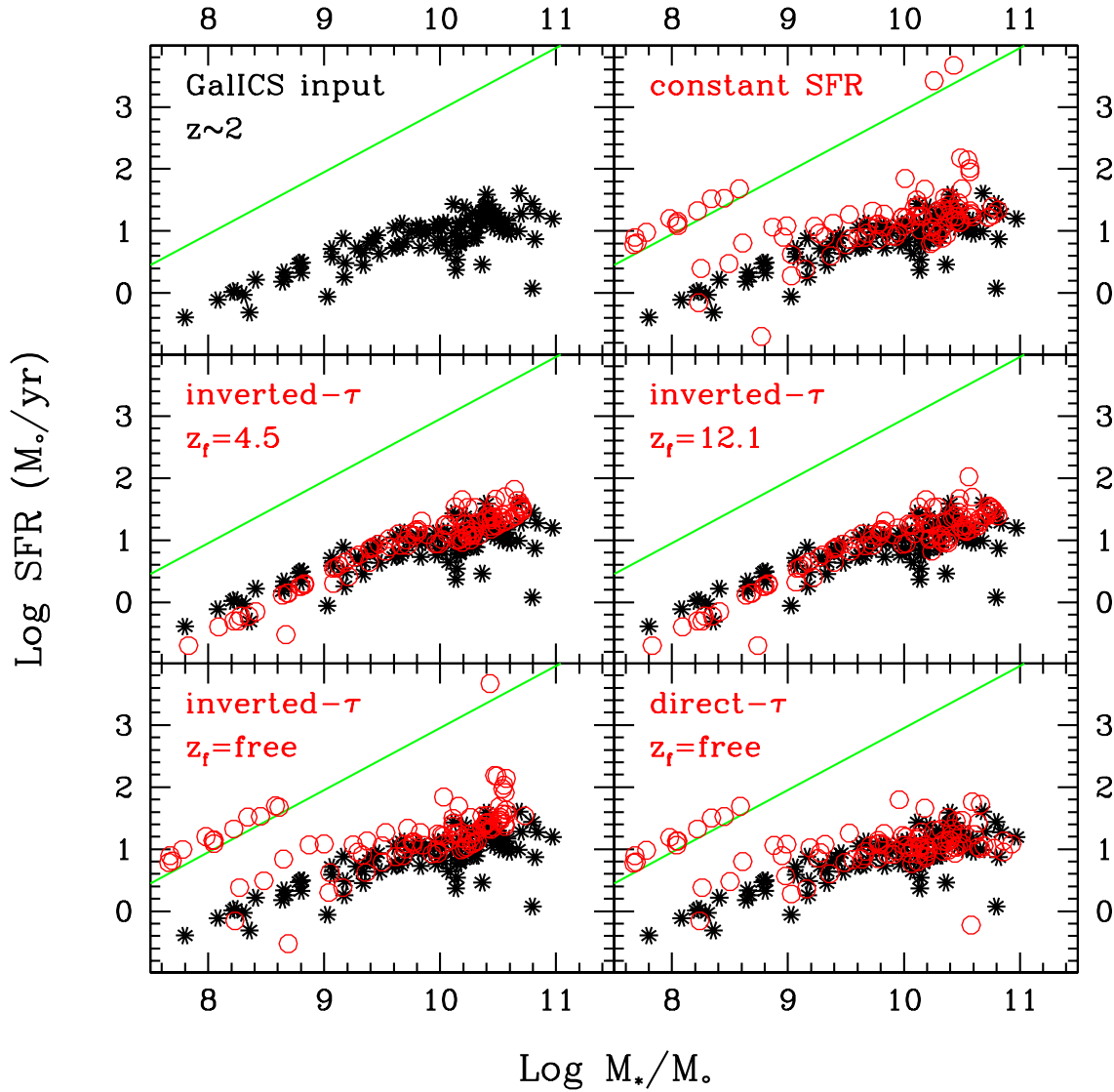}
\caption{Comparison between the input values of SFR and stellar mass
of Mock galaxies from semi-analytic models (labelled as GALICS, black
points) and the same quantities derived from SED fitting using the
various templates (labelled in each panel, red points). The green line
highlights the position of the fake outliers.\label{mock1}}
\end{figure*}

\begin{figure*}
\includegraphics[width=0.9\textwidth]{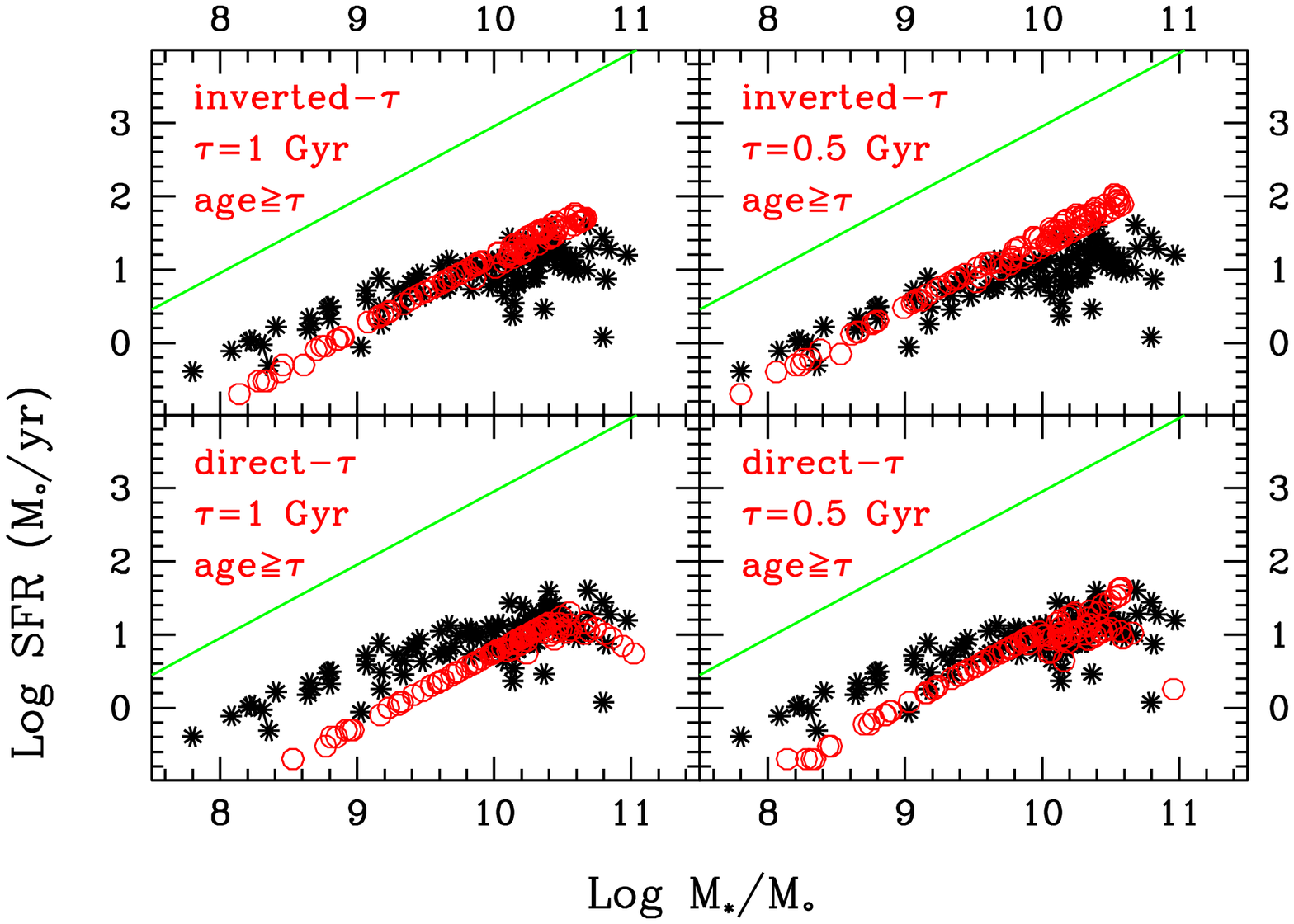}
\caption{The same as Figure~\ref{mock1}, for the cases in which $\tau$ has been fixed to the indicated values, with the additional restriction imposing the age to exceed $\tau$.
\label{mock2}}
\end{figure*}

Figures~\ref{mock1} and \ref{mock2} show the results. The upper left
panel of Figure~\ref{mock1} shows the {\it input} values, i.e. the
distribution of SFRs and masses of the semi-analytic models at
redshift 2 (black points). The other panels show the {\it output}
values, i.e. the SFRs and masses that are obtained by fitting the SED
of the mock galaxies with the various templates, as indicated, over-plotted to the input values.

The inverted-$\tau$ models with fixed formation redshift can recover the input
quantities strikingly well (middle panels). Towards the low-mass end
($M$\lapprox $10^{9} M_{\odot}$) the SFR is somewhat underestimated,
suggesting that the assumed start of SF is too early. 
 Note also, as shown by the two middle panels in
Figure~\ref{mock1}, that the result is almost independent of the
assumed formation redshift, as already argued in Section 4.2.

None of the other SFH templates do as well. With the free age mode a substantial number of outliers with very high SFRs are obtained.  The green line in Figure~\ref{mock1} and Figure~\ref{mock2} is a fit to the outliers found among the real GOODS galaxies from the direct-$\tau$ models (the blue objects in Figure~\ref{sfrcomp} and Figure~\ref{ebvcomp}), and it is quite instructive to note that when the fit produces wrong values, those follow the same relation of fake values in mock galaxies.  Finally, Figure~\ref{mock2}, analogous to Figure ~\ref{mock1}, shows the result of the same experiment with mock galaxies, but now using the template SFHs discussed in the previous subsection, i.e., with fixed $\tau$ and age $>\tau$. The inverted-$\tau$ models with age $\ge\tau$, and $\tau=0.5$ Gyr give  fairly good fits to the actual values in the mock galaxies, but not as good as those with a prior on the formation redshift (see Figure~\ref{mock1}, middle panels).  Finally, note that some outliers among the mock galaxies, a few with low SFR for their masses (hence almost passive), are poorly reproduced by the explored SFHs, as all assume that star formation is still going on.


The reason why inverted-$\tau$ models fit better is very simple. Most mock galaxies constructed from semi-analytic models exhibit secularly increasing SFRs, as do the cold-stream
hydrodynamical models of Dekel et al. (2009) as one expects from the empirical SFR-mass relation
(e.g., Renzini 2009).  Therefore, we believe that quite enough arguments militate in favour of preferring
exponentially increasing SFRs, as opposed to decreasing or constant. 

\subsection{Allowing very small values of $\tau$}

\begin{figure*}
\includegraphics[width=0.8\textwidth]{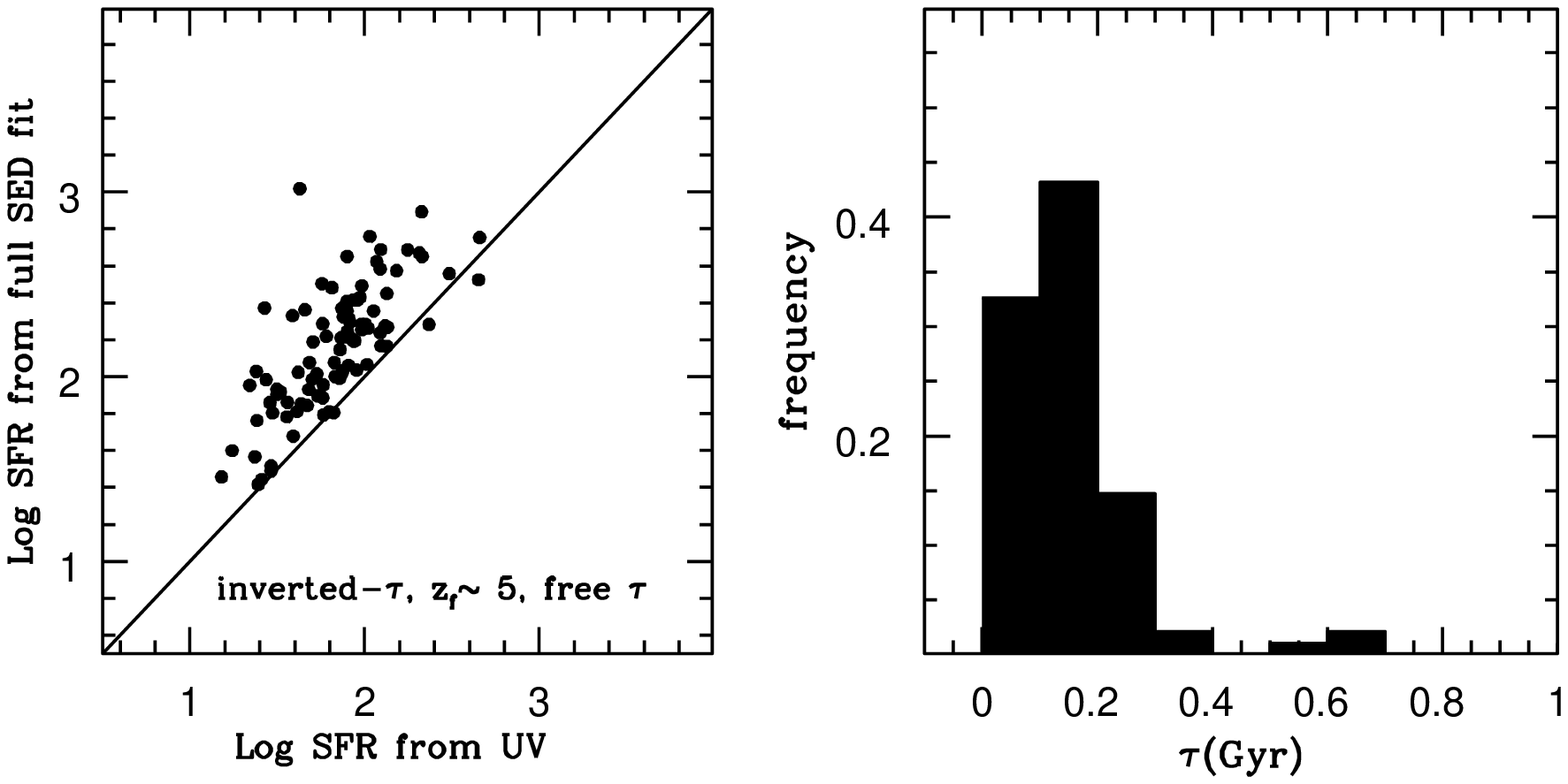}
\caption{Left panel: The SFR from the full optical-to-near-IR SED fit derived from inverted-$\tau$ models in which $\tau$ is allowed to
take very small values, compared to the SFR derived only from the UV part of the SED. Right panel: The frequency histogram of the corresponding values of $\tau$. \label{smalltau}}
\end{figure*}

As one can notice from Figure~\ref{tauzf5}, using inverted-$\tau$ models the vast majority of the best fits tend to cluster close to the minimum allowed value of $\tau$, i.e., 0.3 Gyr, a limit that was imposed based on the value of $\tau$ implied by the empirical SFR$-M_*$ relation, as discussed in Section 4.4. But what happens if this limit is removed, and the best fit procedure is allowed to chose any value of $\tau$ down to 0.05 Gyr? The results are shown in Figure~\ref{smalltau}: several galaxies are now ``best fitted''  with very small $\tau$'s, implying SFHs in which most stars formed just shortly before the time at which the galaxy is observed. This is to say that leaving the procedure free to choose very small values of $\tau$ results  in best fits that closely resemble those obtained with age as a free parameter, that were discussed (and rejected) in Section 4.1. Figure~\ref{smalltau}  shows that also in this case SFRs well in excess of those derived from the UV are derived for several galaxies. Therefore, also inverted-$\tau$ models with unconstrained $\tau$'s suffer from the outshining phenomenon that
plagues models with age as a fully unconstrained free parameter.

The lesson we learn from these experiments is that only by setting a prior on age one derives astrophysically more acceptable solutions. Depending on the adopted shape of the SFH, this prior can be a minimum age (for constant SFR models), or setting age$>\tau$ (for direct-$\tau$ models), or finally setting a minimum acceptable value for $\tau$ (for inverted-$\tau$ models). This kind of fix looks quite artificial, and admittedly the resulting procedure is far from being elegant. It is a simple way of avoiding the outshining effect, and obtain SFRs that are in agreement with those derived from other direct methods, such as from UV, radio,  or mid-IR (e.g., Daddi et al. 2007a,b; Pannella et al 2009).

\subsection{How robust are stellar mass estimates?}

While SFR and reddening can be best evaluated by just using the UV part of a galaxy's spectrum, the determination of the stellar mass, such a crucial quantity for understanding galaxy evolution, requires the full SED fitting. Hence, it requires exploring a variety of assumed SFHs. We have shown that there are some adopted SFHs that, constrained to avoid the outshining problem, give SFRs and $E(B-V)$ values in excellent agreement with those derived from the UV. Now, how different are the derived stellar masses 
when using these viable options for the SFH? And, ultimately, how much can we trust such derived masses?

Figure~\ref{masse} - first and second panel - compares the stellar masses that are derived for the real galaxies studied in this paper using these viable SFHs, namely inverted-$\tau$~with high formation redshift,  inverted-$\tau$~with age constrained, and constant SF with age constrained. The corrensponding masses agree with each other quite well, systematic differences are quite small ($\lapprox 0.1-0.2$ dex), and also the scatter is quite modest, of the order of 0.1 dex. Can this consistency suffice to conclude that the adopted procedures give a robust estimate of stellar masses? In this respect, the fairly accurate recovery of the masses of mock galaxies should increase our confidence.

Yet, the problem of outshining remains. What we consider viable options are SFHs described by extremely simple functions, i.e., either constant or exponential. Nature certainly realises far more complex SFHs.
For example, an early burst of star formation, followed by a continuous SFR, could be easily missed  by the best fit procedure, by remaining out-shined by the successive and ongoing star formation. Thus, we suspect that derived masses may underestimate the actual stellar mass, at least for those galaxies with early massive starbursts.

We note that Figure~\ref{masse}, third panel compares also the masses derived from direct-$\tau$ models with unconstrained age, to those derived from the viable SFHs.  Clearly, such direct-$\tau$ models can substantially underestimate the stellar mass,
while they overestimate the SFR (cf. Figure~\ref{sfrcomp}).

To conclude this section, Figure~\ref{massefinali} shows the {\it absolute} values of the stellar mass of star-forming galaxies at redshift $z\sim2$~that are obtained using our preferred star formation history.


\section{Discussion and Conclusions}
We have tried different priors for the SFH of $z\sim 2$ galaxies, and
used a $\chi^2$ best fit approach to derive the basic properties of
such galaxies: SFR, stellar mass, reddening, age, etc. We do not
expect that any of the simple mathematical forms adopted for the SFH
(constant SFR, exponentially decreasing, exponentially increasing) is
strictly followed by real galaxies. We expect, however, that the two
opposite choices for the exponential case are sufficiently extreme to
encompass the real behavior of galaxies, or at least those having
experienced a quasi-steady SFR over most of their lifetime. We try
various tests to distinguish which of the various options is the most
acceptable one (or the least unacceptable) for giving the
astrophysically most plausible values for the basic properties of star
forming galaxies in the redshift range between $\sim 1.4$ and $\sim
2.5$.

We find that by leaving age as a free parameter the best fit procedure
delivers implausibly short ages, no matter which SFH template is
adopted. This is so because the SED is dominated by the stars that
have formed most recently, and they outshine the older stellar
populations that may inhabit these galaxies. Thus, introducing a prior
on the beginning of star formation appears to be necessary.

Assuming that star formation started at high redshift (i.e., $z>>\sim
2$, the precise value being almost irrelevant) provides a more
credible framework, with models with exponentially increasing SFR
(that we call inverted-$\tau$ models) giving systematically higher
SFRs for the $z\sim 2$ galaxies in the test sample, and lower stellar
masses, compared to models with exponentially decreasing SFR. These
two systematic differences together imply a specific SFR that is a
factor of $\sim 5$ higher in inverted-$\tau$ models, compared to
direct-$\tau$ models, when star formation is assumed to start at high redshift (e.g., $z=5$) 
in both cases.

\begin{figure*}
\includegraphics[width=0.8\textwidth]{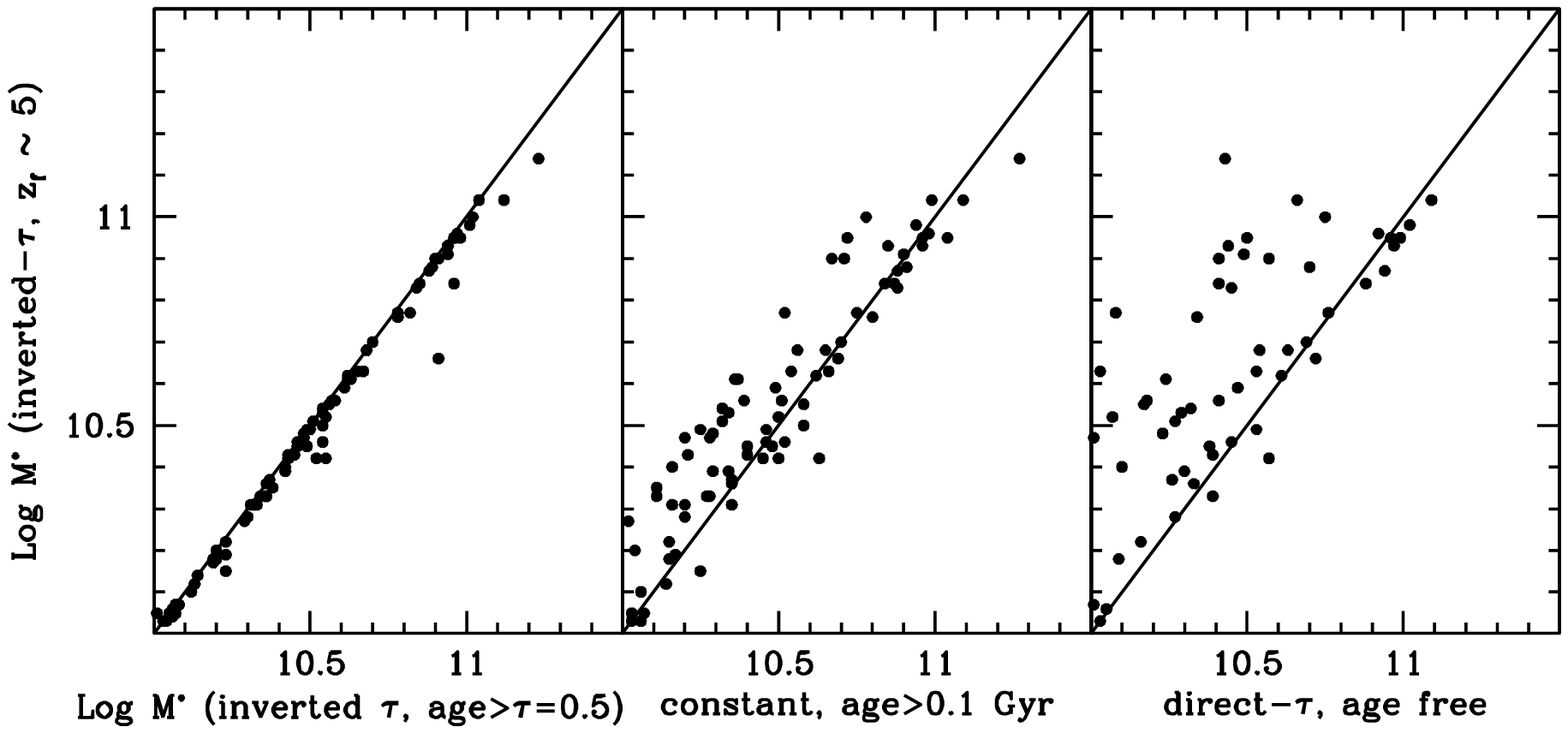}
\caption{A comparison of the stellar masses derived for the GOODS galaxies from inverted-$\tau$ models, with $\zform=5$ 
and $\tau\ge 0.3$ Gyr,  with those derived from other adopted SFHs. Left panel: vs. inverted-$\tau$ models with age $>\tau=0.5$ Gyr; Central panel: vs. models with SFR=const. and age$>0.1$ Gyr; Right panel: vs. direct-$\tau$ models and unconstrained age.
\label{masse}}
\end{figure*}

On the other hand, inverted-$\tau$ models offer various notable
advantages: 1) they indicate SFRs and extinctions in excellent
agreement with those derived from the rest-frame UV part of the SED,
which most directly relates to the ongoing star formation and
reddening, 2) they fairly accurately recover the SFRs and masses of
mock galaxies constructed from semi-analytic models, in which SFR are indeed secularly increasing in most cases, and 3) have
systematically better reduced $\chi^2$ values compared to models with
exponentially decreasing SFRs, though one cannot consider them fully
satisfactory. These advantages are somewhat reduced if the best fit procedure is 
allowed to pick very small values of $\tau$, down to 0.05 Gyr, as in such case
derived SFRs tend to be somewhat overestimated compared to those derived from the UV.

We have also explored star formation histories (such as constant SFR, or exponentially decreasing) in which age is a priori constrained to avoid very small values (e.g., age $>0.1$ Gyr, or age $>\tau$, or fixed $\tau$).  Adopting such priors, though it may seem quite artificial, gives more plausible
solutions with respect to cases in which age is left fully free. This is due to the mentioned outshining effect
of very young stellar populations over the older ones. Still, models with exponentially increasing SFRs tend to give systematically better results, as judged from the consistency with the UV derived SFRs.

Besides the above arguments, there are other reasons to prefer stellar
population models constructed with exponentially increasing SFRs,
including: \par\noindent 1) The almost linear relation between SFR and
stellar mass (SFR $\propto\;\sim M_*$), and the small scatter about
it, that are empirically established for redshift $\sim 2$ galaxies
implies that they can indeed experience a quasi-exponential growth at
these cosmic epochs. Later, some process may quench their star
formation entirely and turn them into passively evolving {\it
ellipticals}, or keep growing until the secular decrease of the
specific star formation rate (the $t^{-2.5}$ term in Equation
(1)) takes over, and galaxies (such as spirals) continue forming
stars at a slowly decreasing rate all the way to the present epoch
(Renzini 2009).

\par\noindent 2) The direct
observation of the evolution of the luminosity function in rest-frame
UV (which is directly related to SFR) shows that the characteristic
luminosity at 1600 \AA\ ($M^*_{1600}$) brightens by over one magnitude
between $z=6$ and $z=3$ (Bouwens et al. 2007).  Since extinction is
likely to increase with time following metal enrichment, the increase
of the rest-frame UV luminosity can only be due to an increased SFR in
individual galaxies.

The aim of this paper is to explore which assumptions and procedures
are likely to give the most robust estimates of the basic stellar
population parameters, when using data that extend from the
rest-frame UV to the near-IR. We argue that ongoing SFRs are best derived at wavelengths below about one micron by using the UV part of the SED of galaxies, and inverted-$\tau$ models give SFRs in excellent agreement with
them. However, the fit to the full optical to near-infrared SED is required to derive the mass
in stars. We show that the SFHs, with a prior on age or $\tau$ to
reduce the outshining effect, give results that agree quite well with
each other, giving some confidence on the reliability of the derived
masses. However, it is still possible that the mass may be somewhat
underestimated for those galaxies in which a substantial fraction of
the stellar mass was produced in a strong burst, early in their
evolution.

\begin{figure}
\includegraphics[width=0.49\textwidth]{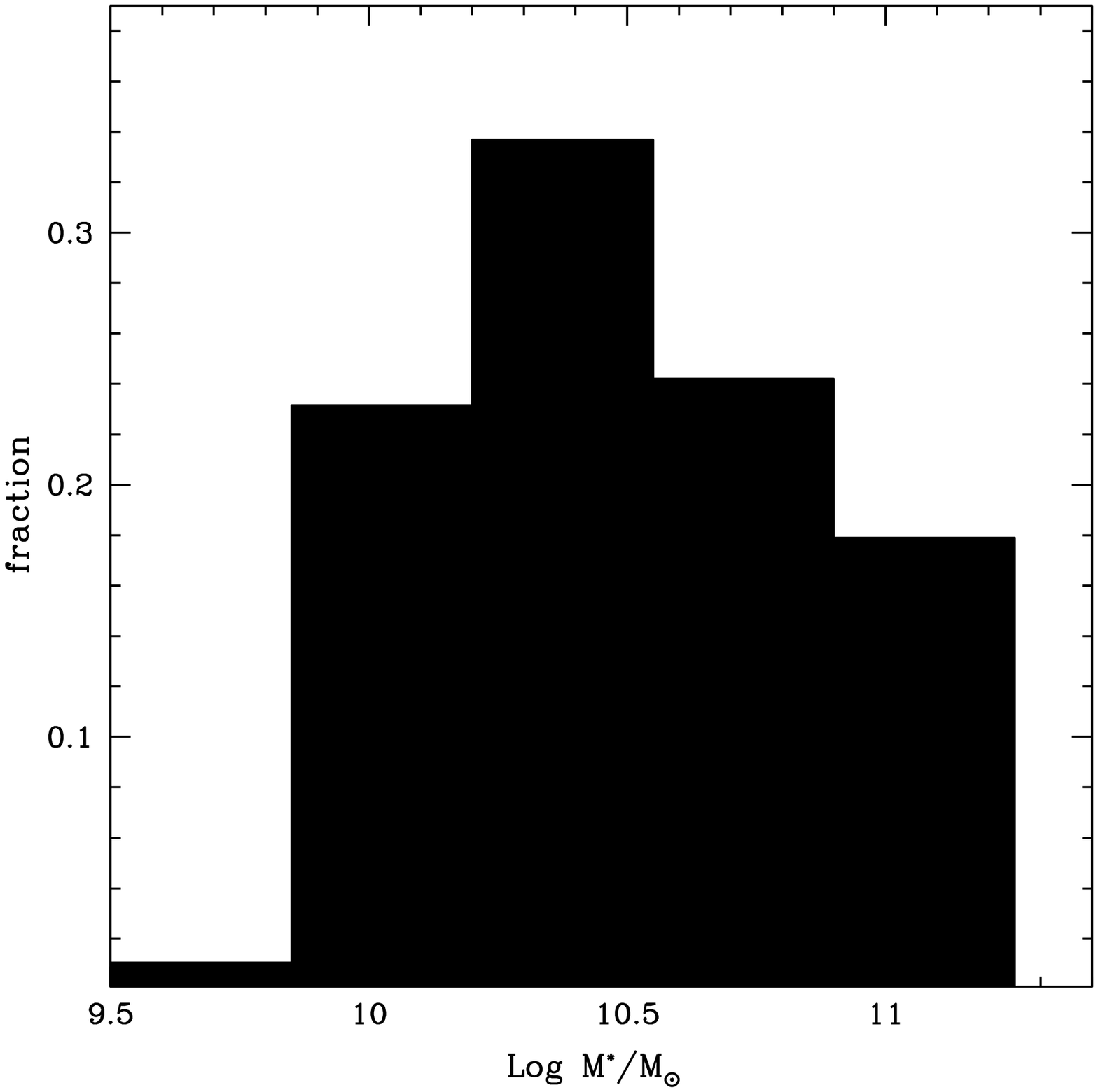}
\caption{Stellar masses for the $z\sim2$~GOODS star-forming galaxies as derived from inverted-$\tau$ models, with 
$\zform\gapprox5$ and $\tau\ge 0.3$~Gyr.
\label{massefinali}}
\end{figure}

At first sight the higher SFRs and lower stellar masses indicated by
inverted-$\tau$ models (compared to direct-$\tau$ models, both with
fixed formation redshift) may exacerbate an apparent mismatch between
the {\it cosmic} (volume averaged) SFR history ($\dot{\rho_*}(z)$) and the
empirical stellar mass density ($\rho_*(z)$), such that the
integration of SFR$(z)$ tends to overproduce the stellar mass density
at most/all redshifts (e.g., Wilkins, Trentham \& Hopkins
2008). However, SFRs are more often derived from direct SFR
indicators, such as the rest-frame UV, radio flux, H$\alpha$ flux, or
mid/far-infrared, rather than from SED fits. Therefore the 0.5 dex
effect on the SFRs relative to the direct-$\tau$ models should not
affect this discrepancy. There may remain the 0.2 dex
effect on the masses, but inverted-$\tau$ models, by construction
minimising star formation at early times, may systematically
underestimate the stellar mass of galaxies, in particular if part of
it was formed in early bursts.  Thus, with this caveat in mind, the
use of inverted-$\tau$ models with fixed formation redshift should not
appreciably exacerbate the mismatch problem mentioned above. Instead,
direct-$\tau$ models with unconstrained age most certainly worsen this problem as
they underestimate the stellar mass (see Figure~\ref{masse}, right
panel), and overestimate the SFR (see Figure~\ref{sfrcomp}).

We conclude that the use of synthetic stellar populations with an
exponentially declining SFR should be avoided in the case of star forming galaxies
at redshift beyond $\sim 1$. Exponentially increasing SFRs, with
e-folding times of $\sim 0.3-1$ Gyr, and a high starting redshift appear to provide astrophysically
more plausible results, and while nature may not closely follow such a
simple functional behavior, they should be preferred in deriving SFRs
and stellar masses of high redshift galaxies. In any event, we believe that 
maximum likelihood parameters are not necessarily unbiased estimators of star formation rates and masses of high-redshift galaxies, but the broadest possible 
astrophysical context should be considered.

\section*{Acknowledgments}
We thank the referee Stephen Serjeant for a prompt and useful report.
We are grateful to Andi Burkert for valuable comments on a very early draft 
of the paper, and to Laura Greggio and Daniel Thomas for useful discussions. 
We acknowledge Micol
Bolzonella for her constructive help with the {\it Hyper-Z} code.  CM, JP, and CT
acknowledge the Marie-Curie Excellence Team grant "Unimass", ref.
MEXT-CT-2006-042754 of the Training and Mobility of Researchers
programme financed by the European Community. ED acknowledges the funding 
support of  the ERC-StG grant UPGAL-240039,
ANR-07-BLAN-0228 and ANR-08-JCJC-0008. AR is grateful for the hospitality of the ICG during the final rush for
the completion of this paper.
Some of the data used
here are part of the GOODS Spitzer Space Telescope Legacy Science
Program, that is supported by NASA through Contract Number 1224666
issued by the JPL, Caltech, under NASA contract 1407.

\end{document}